\PassOptionsToPackage{hyphens}{url}
\documentclass[10pt,journal,compsoc]{IEEEtran}
\ifCLASSOPTIONcompsoc
\usepackage[nocompress]{cite}
\else
\usepackage{cite}
\fi

\ifCLASSINFOpdf
\else
\fi
\usepackage[numbers]{natbib}
\usepackage{hyperref}
\hypersetup{
colorlinks = true, % false: boxed links; true: colored links
linkcolor=black, % color of internal links
citecolor=black, % color of links to bibliography
urlcolor=black % color of external links
}

\usepackage[linesnumbered,ruled,vlined]{algorithm2e}
\usepackage{subcaption}

\usepackage{graphicx}
\usepackage{float} 
\usepackage{caption}

\usepackage{stfloats}
\usepackage{comment}
\usepackage{makecell}
\usepackage{graphicx}
\usepackage{colortbl}
\usepackage{paralist}
\usepackage{color}
\usepackage{booktabs} % For formal tables
\usepackage{xspace}
\usepackage{pifont}
\usepackage{amsmath}
\usepackage{amsthm}
\usepackage{amssymb}
\usepackage{tikz}
\usepackage{multirow}
\usepackage{threeparttable}
\usetikzlibrary{automata,positioning}
\usepackage{pgfplots,pgfplotstable}
\usepackage[breakable]{tcolorbox}
\usepgfplotslibrary{statistics}
\usetikzlibrary{arrows,patterns}
\pgfplotsset{
discard if not/.style 2 args={
x filter/.append code={
\edef\tempa{\thisrow{#1}}
\edef\tempb{#2}
\ifx\tempa\tempb
\else

\fi
}
}
}

\pgfplotsset{compat=newest}

\makeatletter \newcommand{\pgfplotsdrawaxis}{\pgfplots@draw@axis} \makeatother

\pgfplotsset{axis line on top/.style={
axis line style=transparent,
ticklabel style=transparent,
tick style=transparent,
axis on top=false,
after end axis/.append code={
\pgfplotsset{axis line style=opaque,
ticklabel style=opaque,
tick style=opaque,
grid=none}
\pgfplotsdrawaxis}
}
}

\pgfplotsset{
discardbp if not/.style 2 args={
/pgfplots/boxplot/data filter/.append code={
\edef\tempa{\thisrow{#1}}
\edef\tempb{#2}
\ifx\tempa\tempb
\else

\fi
}
}
}

\theoremstyle{definition}

\providecommand{\definitionname}{Definition}

\theoremstyle{definition}

\providecommand{\examplename}{Example}

\newcommand{\method}{\ensuremath{\textsc{SafeVar}}\xspace}
\newcommand{\VehicleCharacteristicsSetting}{VCS\xspace}
\newcommand{\VehicleCharacteristicsSettings}{VCSs\xspace}
\newcommand{\OriVehicleCharacteristicsSetting}{VCS$_{orig}$\xspace}
\newcommand{\OriVehicleCharacteristicsSettings}{VCSs$_{orig}$\xspace}
\newcommand{\categoryOfcharacteristicsChanged}{CC}

\newcommand{\maxRpm}{\texttt{max\_rpm}\xspace}
\newcommand{\mass}{\texttt{mass}\xspace}
\newcommand{\radius}{\texttt{radius}\xspace}
\newcommand{\dampingRateFullThrottle}{\texttt{dampRateFullT}\xspace}
\newcommand{\dampingRateZeroThrottleClutchEngaged}{\texttt{dampRateZeroT\_CE}\xspace}
\newcommand{\dampingRateZeroThrottleClutchDisengaged}{\texttt{dampRate\_zeroT\_CD}\xspace}
\newcommand{\gearSwitchTime}{\texttt{gearSwitchTime}\xspace}
\newcommand{\clutchStrength}{\texttt{clutchStrength}\xspace}
\newcommand{\dragCoefficient}{\texttt{dragCoeff}\xspace}
\newcommand{\tireFriction}{\texttt{tireFric}\xspace}
\newcommand{\dampingRate}{\texttt{dampRate}\xspace}
\newcommand{\maxBrakeTorque}{\texttt{maxBrakeTorque}\xspace}

\newcommand{\car}{\ensuremath{\mathtt{VUT}}\xspace}
\newcommand{\carp}{\ensuremath{\car_{\params}}\xspace}

\newcommand{\carv}{\ensuremath{\car_{\origParamValues}}\xspace}
\newcommand{\origParamValues}{\ensuremath{\overline{v}}\xspace}
\newcommand{\modParamValues}{\ensuremath{\origParamValues^\prime}\xspace}
\newcommand{\filteredModParamValues}{\ensuremath{\origParamValues^{\prime\prime}}\xspace}
\newcommand{\carvp}{\ensuremath{\car_{\modParamValues}}\xspace}
\newcommand{\carvpp}{\ensuremath{\car_{\filteredModParamValues}}\xspace}
\newcommand{\paramsSet}{\ensuremath{\mathit{Characs}}\xspace}
\newcommand{\params}{\ensuremath{\overline{c}}\xspace}

\newcommand{\scen}{\ensuremath{S}\xspace}
\newcommand{\scenSearch}{\ensuremath{\Tilde{\scen}}\xspace}
\newcommand{\simulation}{\ensuremath{\mathtt{simulation}}\xspace}
\newcommand{\res}{\ensuremath{\mathit{res}}\xspace}
\newcommand{\safetyDegree}{\ensuremath{\mathit{safetyDegree}}\xspace}
\newcommand{\filterV}{\ensuremath{\mathit{filter}}\xspace}
\newcommand{\fitSaf}{\ensuremath{\mathit{f_{safe}}}\xspace}
\newcommand{\fitDiff}{\ensuremath{\mathit{f_{diff}}}\xspace}
\newcommand{\fitNumDiff}{\ensuremath{\mathit{f_{numDiff}}}\xspace}
\newcommand{\searchVars}{\ensuremath{\overline{x}}\xspace}
\newcommand{\thresholdI}{\ensuremath{\mathit{Th}_i}\xspace}

\newcommand{\paretoFront}{\ensuremath{\mathit{PF}}\xspace}

\newcommand{\Atwelve}{\ensuremath{\hat{A}}\textsubscript{12}\xspace}
\newcommand{\precision}{\ensuremath{\beta}\xspace}

\newcommand{\carla}{CARLA\xspace}

\newcommand{\valueDiff}{\ensuremath{\Delta}\xspace}
\newcommand{\percChange}{PC\xspace}
\newcommand{\wheelMass}{\texttt{wheel\_mass}\xspace}

\newcommand{\minRpm}{\texttt{min\_rpm}\xspace}
\newcommand{\maxMotorTorque}{\texttt{maxMotorTorque}\xspace}
\newcommand{\maxSteeringAngle}{\texttt{maxSteeringAngle}\xspace}
\newcommand{\tireDragCoeff}{\texttt{tireDragCoeff}\xspace}
\newcommand{\wheelDamping}{\texttt{wheel\_damping}\xspace}
\newcommand{\shiftTime}{\texttt{shiftTime}\xspace}
\newcommand{\tractionControlSlipLimit}{\texttt{tractionControlSlipLimit}\xspace}
\newcommand{\precSetCarlaSun}{\textit{Carla Sun}\xspace}
\newcommand{\precSetCarlaRain}{\textit{Carla Rain}\xspace}
\newcommand{\precSetLgsvl}{\textit{LGSVL}\xspace}
\newcommand{\paretoFrontCase}{\ensuremath{\mathit{PF_{th(safe)}}}\xspace}
\newcommand{\thresholdMin}{\ensuremath{\mathit{Th}_{safe}}\xspace}
\newcommand{\lgsvl}{LGSVL\xspace}
\newcommand{\ttc}{\ensuremath{\mathit{TTC}}\xspace}
\newcommand{\ttcStar}{\ensuremath{\mathit{TTC^{*}}}\xspace}
\newcommand{\ttcCritical}{\ensuremath{\mathit{TTC}_{\mathit{critical}}}\xspace}
\newcommand{\tet}{\ensuremath{\mathit{TET}}\xspace}
\newcommand{\tetStar}{\ensuremath{\mathit{TET}^{*}}\xspace}
\newcommand{\tit}{\ensuremath{\mathit{TIT}}\xspace}
\newcommand{\titStar}{\ensuremath{\mathit{TIT}^{*}}\xspace}
\newcommand{\aveDece}{\ensuremath{\mathit{aveDece}}\xspace}
\newcommand{\aveSafetyDegreeD}{\ensuremath{\mathit{avgSD_{mD}}}\xspace}
\newcommand{\aveSafetyDegreeCS}{\ensuremath{\mathit{avgSD_{cS}}}\xspace}
\newcommand{\velocity}{\ensuremath{v_{L}}\xspace}
\newcommand{\relativeDistance}{\ensuremath{L_{s}}\xspace}
\newcommand{\timeStep}{\ensuremath{\tau_{\mathit{sc}}}\xspace}

\newcommand{\fuzzerTestCase}{\ensuremath{\mathit{VCS_{unsafe}}}\xspace}

\newcommand{\safeFuzzer}{\ensuremath{\mathit{SafeFuzzer}}\xspace}

\newcommand{\fitnessScore}{\ensuremath{\mathit{Fitness\ Score}}\xspace}

\usepackage{todonotes}
\begin{document}
%\title{Search-Based the Analysis of the Impact of Vehicle Configuration Variations on the Autonomous Driving Systems}

\title{Safety Assessment of Vehicle Characteristics Variations in Autonomous Driving Systems}
% received ..." text while in non-compsoc journals this is reversed. Sigh.

\author{Qi Pan,
Tiexin Wang,
Paolo Arcaini,
Tao Yue,
Shaukat Ali% <-this % stops a space
\IEEEcompsocitemizethanks{
\IEEEcompsocthanksitem T. Yue (the corresponding author) is with the School of Computer Science and Engineering, Beihang University, Beijing, China. E-mail: yuetao@buaa.edu.cn
\IEEEcompsocthanksitem Q. Pan and T. Wang are the College of Computer Science and Technology, Nanjing University of Aeronautics and Astronautics, Nanjing, China. E-mail: \{panq, tiexin.wang\}@nuaa.edu.cn
\IEEEcompsocthanksitem P. Arcaini is with National Institute of Informatics, Tokyo, Japan. E-mail: arcaini@nii.ac.jp.
\IEEEcompsocthanksitem S. Ali is with Simula Research Laboratory and Oslo Metropolitan University, Oslo, Norway. E-mail: shaukat@simula.no.

}% <-this % stops an unwanted space
\thanks{Manuscript received April 19, 2005; revised August 26, 2015.}}

\IEEEtitleabstractindextext{%vehicle
\begin{abstract}
Autonomous driving systems (ADSs) must be sufficiently tested to ensure their safety. Though various ADS testing methods have shown promising results, they are limited to a fixed set of vehicle characteristics settings (\VehicleCharacteristicsSettings). The impact of variations in vehicle characteristics (e.g., mass, tire friction) on the safety of ADSs has not been sufficiently and systematically studied. Such variations are often due to wear and tear, production errors, etc., which may lead to unexpected driving behaviours of ADSs. To this end, in this paper, we propose a method, named \method, to systematically find minimum variations to the original vehicle characteristics setting, which affect the safety of the ADS deployed on the vehicle. To evaluate the effectiveness of \method, we employed two ADSs and conducted experiments with two driving simulators. Results show that \method, equipped with NSGA-II, generates more critical \VehicleCharacteristicsSettings that put the vehicle into unsafe situations, {as compared with two baseline algorithms: Random Search and a mutation-based fuzzer. We also identified critical vehicle characteristics and reported to which extent varying their settings put the ADS vehicles in unsafe situations.}

%Furthermore, we studied the impact of weather conditions. Experiment results show that slight variations to some characteristics under certain weather conditions (e.g., rain) might put ADSs into unsafe situations such as collisions.
%, demonstrated the importance of studying the effects of multiple parameter configurations and their interactions in a single search on collision, and analyzed the differences in vehicle configuration variations under different weather conditions.
\end{abstract}

}

\maketitle

\IEEEdisplaynontitleabstractindextext

% \fi
\IEEEpeerreviewmaketitle

\IEEEraisesectionheading{\section{Introduction}\label{sec:introduction}}


% focus on ADS not only ADAS (background)
\IEEEPARstart{A}utomated Driving Systems (ADSs) are being developed to bring many benefits, such as significantly reducing accidents and congestion~\cite{nstc2020ensuring}. However, ADSs are complex since they process continuous heterogeneous data, thereby implementing complicated functional logic and increasingly using AI algorithms. Consequently, testing them to ensure their safety is a big challenge nowadays.

Various ADS testing techniques have been proposed, {including methods detailed in the works of, for example, Abdessalem et al.~\cite{Abdessalem2016ASE,Abdessalem2018TAC,abdessalem2018TVC}, Gladisch et al.~\cite{ASESBTADS}, Li et al.~\cite{LiISSRE20}, Gambi et al.~\cite{gambi2019automatically}, Liu et al.~\cite{reqsViolADSsASE21}, Cal\`{o} et al.~\cite{avoidCollICST2020,calo2020simultaneously}; please refer to Tang et al.~\cite{TangTOSEM2023} for a recent survey}. Most of them focus on scenario-based testing with {simulators~\cite{zhong2021survey}}, which intend to identify critical driving or test scenarios cost-effectively. Such scenarios often involve meteorological elements (e.g., wind, temperature, pressure, humidity, clouds, and precipitation), traffic environment elements (e.g., physical condition and geometry of the road surface, signs), and driving environment elements (e.g., nonplayer characters (NPC) vehicles, pedestrians), etc. These approaches rarely consider variations in the setting of vehicle characteristics (e.g., mass, tire friction, and radius) due to production variation, wear and tear, etc. However, as reported {by Lee et al.~\cite{lee2019stability} and Zhang et al.~\cite{zhang2020investigating,incrApprPathPlannerTDSC2022}}, such variations may significantly impact the safety of vehicle behaviours. %For instance, minor variations in a \VehicleCharacteristicsSetting, such as tires or brake pad wear, may cause unsafe ADS behaviours. 
%The abnormal behaviours (e.g., unintended acceleration, unknown swing) may consequently pose severe threats to the safety of the autonomous vehicle.
Hence, it is essential to study vehicle characteristics and their interactions that negatively impact ADS safety and the level of the impact. 
To this end, we propose \method, aiming to find minimum variations to \emph{original vehicle characteristics settings}, (i.e., \OriVehicleCharacteristicsSettings) such that the safety of the ADS under study is decreased. \OriVehicleCharacteristicsSettings of a virtual vehicle are provided by the default settings of the simulator on which the virtual vehicle runs.
%, and ADSs are trained based on such vehicle dynamics model (under original characteristic values), such as Apollo\footnote{https://github.com/ApolloAuto/apollo} employs different types of vehicles to test, the algorithm of Apollo would call different profiles depending on the type of vehicle.
{The variations identified by \method could be used to improve the ADS and the vehicle designs and guide test engineers to configure critical characteristics of virtual vehicles when performing simulation-based testing of ADSs.}

Specifically, \method relies on simulators (e.g., \carla~\cite{CARLA}) and employs a multi-objective approach (with the search algorithm NSGA-II) to search for vehicle characteristics settings (\VehicleCharacteristicsSettings) such that minimum variations to \OriVehicleCharacteristicsSettings lead to the highest impact on the safety of the ADS; the search has three optimisation objectives: 
\begin{inparaenum}[1)]
\item minimising the safety of the ADS under study, 
\item minimising the variations to \OriVehicleCharacteristicsSettings, and
\item minimising the number of characteristics to change.
\end{inparaenum}

% experiment design

In our experiments, we employed \carla~\cite{CARLA} and \lgsvl~\cite{rong2020lgsvl} as the simulators to simulate driving scenarios, two ADSs (i.e., World On Rails~\cite{WOR} and Apollo\footnote{https://github.com/ApolloAuto/apollo}) and two virtual vehicles corresponding to the real vehicle 2017 Lincoln MKZ. {We acknowledge the existence of various simulators, such as BeamNG.Tech \cite{beamNG} and Gazebo \cite{koenig2004design}. The selection of CARLA and LGSVL is due to their open-source and high-fidelity nature and easy integration with existing systems such as Apollo.}
A set of 12 vehicle characteristics characterises each virtual vehicle. With \carla, we experimented with two weather conditions (sunny and rainy days). To evaluate the use of NSGA-II in \method, we employ Random Search (RS) {and a mutation-based fuzzer (called \safeFuzzer in the following) as the baselines}. {Results show that NSGA-II significantly outperformed RS and \safeFuzzer regarding Inverted Generational Distance (IGD)}, i.e., a commonly used quality indicator in search, and four safety metrics (e.g., the safety degree). We also observed that NSGA-II generates \VehicleCharacteristicsSettings that have a higher chance of putting the vehicle into unsafe situations than its \OriVehicleCharacteristicsSetting, and these \VehicleCharacteristicsSettings often involve value changes of 4-6 characteristics with \mass and \maxBrakeTorque as the most frequently changed characteristics.  
In our previous work~\cite{yin}, we proposed a multi-objective search-based approach to analyse the impact of \VehicleCharacteristicsSetting variations on the safety of Advanced Driver Assistance Systems (ADASs). We extended this work from the following aspects:
\begin{inparaenum}[(i)]
\item We proposed \method to generate variations to vehicle characteristics that threaten the safety of ADSs (not the Automatic Emergency Brake operation in~\cite{yin});
\item We designed more realistic driving scenarios (e.g., pedestrians crossing the road and avoiding stationary vehicles ahead);
\item With \carla, we conducted experiments considering two different weather conditions; 
\item We experimented \method with two ADSs (World On Rails and Apollo) and two simulators;
\item To more comprehensively evaluate \method, we introduced more safety metrics.% in addition to the safety metric \safetyDegree that we use to drive the search (i.e., Time Exposed Time-to-collision (\tet), Time Integrated Time-to-collision (\tit), and average deceleration (\aveDece)).
\end{inparaenum}

\textit{Paper Structure.} Section~\ref{sec:relatedWork} presents the related work. Section~\ref{sec:approach} introduces \method. We present the empirical study in Section~\ref{sec:evaluation}, followed by discussions (Section~\ref{sec:discussion}), the threats to validity (Section~\ref{sec:threats}), and the conclusion (Section~\ref{sec:conlusion}).

\section{Related Work}\label{sec:relatedWork}
% \IEEEraisesectionheading{\section{Related Work}\label{sec:related work}}
%
%Search-based testing approaches have been proposed and applied for both testing automotive systems including ADASs and ADSs~\cite{Abdessalem2016ASE, Abdessalem2018TAC, Abdessalem2018TVC}. 
{We present the related work from four perspectives:
\begin{inparaenum}[1)]
\item vehicle characteristics and safety variations,
\item testing of ADSs, and
\item multi-objective optimisation in automotive.
\end{inparaenum}
}
\subsection{Vehicle characteristics and safety variations} 
Stellet et al.~\cite{stellet2016analytical} extended the work by Nilsson et al.~\cite{2015Worst} for analysing the impact of noisy sensor measurements and uncertain prediction models (e.g., measurement errors and incomplete environmental perception) on Automatic Emergency Brake (AEB) systems. That work helps define AEB requirements, perform sensitivity analysis, and tune sensor and system parameters (e.g., sampling time). 
Zhang et al.~\cite{zhang2020investigating,incrApprPathPlannerTDSC2022} proposed an approach based on fault localization to assess the relation between the configuration of a path planner (i.e., an ADS functional module) and the degree of safety (i.e., cause or avoid a collision) that can be achieved by the ADS vehicle during driving. 

As discussed by Lee et al.~\cite{lee2019stability}, the safety of automotive products might be significantly affected by small changes in their vehicle properties when they operate under specific driving scenarios, environmental conditions, etc. Based on this observation, Lee et al.~\cite{lee2019stability} formulated a multi-objective optimisation problem and employed NSGA-II to solve it. However, their approach focuses on looking for pairs of \VehicleCharacteristicsSettings (i.e., a pair of similar settings containing the same set of parameters) in which only one parameter changes in a single search. Our approach \method, instead, allows different numbers of parameters to change in a single search.

%Compared with the above works, \method focuses on the impact of vehicle characteristics variations on ADS safety, while~\cite{stellet2016analytical, zhang2020investigating} focus on sensors and ADSs respectively, and we study the effect of the interactions of different sets of parameters while~\cite{lee2019stability} focuses on single parameter changes within a set of parameters. Considering the evaluation of safety performance, we used more metrics. 
%They are intelligent decision-control software and have the ability to respond safely and flexibly to critical scenarios and unexpected situations, rather than linearly responding to changes in the vehicle configuration. In addition, we constructed more realistic traffic scenarios and analyzed the impact of extreme weather.

{\subsection{ADS testing}\label{ads testing}}
%\todo[inline]{To mention somewhere this survey~\cite{TangTOSEM2023}}
%As discussed in the survey by Tang et al.~\cite{TangTOSEM2023}, ADS testing is challenging due to the high complexity of ADSs and their multidisciplinary nature. The survey includes 181 publications and classifies existing ADS testing methods into two categories: module-level and system-level testing, both of which cover scenario-based testing. 
{Testing ADSs is a critical means to ensure their safety. Ding et al.~\cite{10089194} surveyed critical driving scenario generation algorithms, classified them into three categories: {\it data-driven}, {\it adversarial}, and {\it knowledge-based} generation, and discussed simulators and platforms. Tang et al.~\cite{TangTOSEM2023} also surveyed ADS testing by focusing on both module-level and system-level testing, highlighted the gap between ADS testing in simulators and the real world, and summarised challenges. Below, we discuss related works from three aspects: scenario-based ADS testing, testing neural network components of ADSs, and search-based testing of ADSs and ADASs.}

\subsubsection{Scenario-based ADS testing}
{Scenario-based testing mainly focuses on identifying and defining driving scenarios characterised by road conditions, traffic, weather, etc. Zhong et al.~\cite{zhong2021survey} classified scenario-based testing methods into {\it real-world}, {\it hardware-in-the-loop}, and {\it software-in-the-loop}. Moreover, they provided a generic formulation of scenario-based testing in high-ﬁdelity simulation, summarised and compared commonly applied high-fidelity simulators, and discussed challenges in ADS testing, such as the gap between simulation and real-world testing results. In the rest of this section, we discuss several recently published ADS testing approaches.}

%Masuda et al.~\cite{masuda2018rule} proposed to build models capturing the ego vehicle and its interactions with NPC vehicles. Each model is defined as a set of rules specifying and constraining road types, paths, physical road conditions (e.g., the number of lanes), NPC vehicles' speed and position, etc. All these pre-defined rules make up a simulation model. An algorithm was proposed to search collision cases within manually constructed simulation models. %a test case generation approach was proposed, along with a simulation model describing rules between the ADS under test and NPC vehicles, and an algorithm searching test cases that lead to collisions. In this way, the method of defining test cases based on the logical rules of the simulation model can effectively generate test cases for some specific test objectives (i.e., collision test cases). %The results from experiments indicate that the algorithms are superior to those of the baseline method.

Wachi et al.~\cite{wachi2019failure} proposed \textit{FAILMAKER-ADVRL} to train NPC vehicles using multi-agent reinforcement learning (RL) to fail rule-based agents (algorithms under test) and consequently identify failure scenarios. Baumann et al.~\cite{baumann2021automatic} proposed an RL-based optimisation to generate critical scenarios with Q-Learning by focusing on the overtaking assistant feature. {Lu et al. proposed \textit{DeepCollision}~\cite{lu2022learning} and \textit{DeepQTest}~\cite{lu2023deepqtest} to learn, using RL, environment configurations having a high chance of revealing abnormal ADS behaviours. Particularly, DeepQTest introduces real-world
weather condition data into the simulated environment to ensure the realism of generated driving scenarios. 
Cheng et al.~\cite{BehAVExplor2023} proposed \textit{BehAVExplore} to explore different behaviours of ADS vehicles and detect safety violations with an unsupervised model characterising these behaviours.}

%Koren et al.~\cite{koren2018adaptive} applied Deep RL to improve the efficiency of adaptive stress testing of ADSs and demonstrated that their approach can find more likely failure scenarios with higher efficiency than the Monte Carlo tree search. 
%
%Compared with pure parameter combination variations, their approach generates six times more critical test cases.
%

Considering the above-mentioned literature, those works focus more on generating (or changing) environmental elements that may interact with the ADS under test rather than generating or varying vehicle characteristics.

%Researchers have also proposed to use of other Machine-Learning (ML) methods for ADS testing. This kind of method mainly includes generating adversarial cases based on ML or generating new test cases from existing test cases to test deep neural networks in ADS.
\subsubsection{{Testing neural networks of ADSs}}
Researchers have proposed various approaches to test deep neural networks used in ADSs. For example, Pei et al.~\cite{DeepXplore} proposed \textit{DeepXplore} to test deep learning-based systems, including ADSs. \textit{DeepXplore} generates new test cases for deep learning-based systems based on existing testing datasets to discover potential wrong behaviours of the systems. Tian et al.~\cite{TianDeepTest2018} proposed \textit{DeepTest}, which focuses on synthesising new driving scene images by applying image transformations and then uses synthesised images to simulate weather conditions, object movements, etc. {Zhang et al.~\cite{zhang2018deeproad} proposed \textit{DeepRoad}, an unsupervised framework to automatically generate driving scenes to test the consistency of DNN-based ASDs under various weather conditions across different scenes. 
Haq et al.~\cite{haq2021can} conducted an empirical study to investigate how offline and online testing of ADS DNNs differ and concluded that offline testing is less effective than online testing, while prediction errors from offline testing can lead to safety violations that can be detected by online testing.
}
%
%Jha et al.~\cite{JhaML2019} proposed a machine learning-based faults injection engine named \textit{DriveFI}. By mining high-risk scenarios, 561 safety-critical faults were found by \textit{DriveFI} within 4 hours. %By contrast, randomized injections conducted over several weeks found no safety-critical faults.
{
To compare with these approaches, \method is search-based and focuses on the performance of ADSs in terms of safety as a whole application while varying vehicle properties.}

\subsubsection{Search-based testing of ADSs and ADASs}
SBT is widely used for testing ADASs and ADSs~\cite{TangTOSEM2023}. 
For instance, Abdessalem et al.~\cite{Abdessalem2016ASE} proposed a design-time (in simulated environments) testing approach to identify critical scenarios, utilising NSGA-II and surrogate models developed with machine learning techniques. %, with the overall aim of mitigating the computational cost of executing physics-based ADAS simulations. 
Evaluation results show that NSGA-II significantly outperforms RS.% and can identify useful critical scenarios. 
Abdessalem et al.~\cite{abdessalem2018TVC} also proposed to combine NSGA-II and decision tree classification models to identify critical scenarios. Further, Abdessalem et al.~\cite{Abdessalem2018TAC} proposed another SBT approach for detecting feature interaction failures (e.g., AEB and Adaptive Cruise Control). %Evaluation results show that their approach effectively identifies more feature interaction failures than two baselines.
Gambi et al.~\cite{gambi2019automatically} proposed combining procedural content generation and genetic algorithms to generate virtual roads for testing lane-keeping components. Luo et al.~\cite{reqsViolADSsASE21} proposed EMOOD, an SBT approach, to generate scenarios to expose different combinations of requirements violations. %The work was evaluated with an industrial ADS under two traffic situations, and results show that EMOOD's performance was better than the three baselines.
{Lu et al.~\cite{LuSSBSE} proposed \textit{SPECTRE}, which applies multi-objective search algorithms to select and prioritize driving scenarios.} 

{AV-FUZZER is an automatic testing framework proposed by Li et al.~\cite{LiISSRE20} for generating safety violation scenarios by searching for perturbations to traffic participants (e.g., NPC vehicles) such that the safety of an ADS vehicle can be minimized with genetic algorithms.} 
{\textit{DriveFuzz} proposed by Kim et al.~\cite{kim2022drivefuzz}, is another fuzzing testing framework that automatically generates and changes high-fidelity test scenarios by mutating driving factors such as weather and invisible puddles. Zhong et al.~\cite{zhong2022neural} proposed a grammar-based fuzzing technique called \textit{AutoFuzz}, which leverages learning-based seed selection and mutation strategies to reveal more unique traffic violations than existing methods.}

{We also want to acknowledge that, competition on SBT of ADSs has been organised in the SBFT (previously SBST) workshop~\cite{SBST2021competition,SBST2022competition,SBST2023competition}, aiming to test the lane-keeping component of an ADS running in the BeamNG.tech simulator\footnote{\url{https://beamng.tech/}}. Castellano et al.~\cite{freneticSBST2021,freneticVSBST2022} and Klikovits et al.\cite{freneticLibSCICO2023} proposed ways to generate value roads (e.g., not too curved) on which the vehicle drives off the lane. Ferdous et al.~\cite{EvoMBTSBST2022,EvoMBTSBST2023,EvoMBTscico2023} proposed an SBT approach for generating tests of road configurations from extended finite state machines. Peltom{\"a}ki et al.~\cite{WOGANSBST2022} and Winsten and Porres~\cite{WOGANSBST2023} proposed a road generator that uses generative adversarial network (GAN) with the Wasserstein distance; the approach tries to maximise the percentage of the car body out of the lane and the maximum distance from the car to the centre of the lane.}

Compared with these works, \method has a different goal: searching for the minimum number of vehicle characteristics (e.g., the radius of a tire) with minimum changes to their values having a high possibility of leading to a reduction of the ADS safety. 

\subsection{Multi-objective optimisation in automotive} 
Hybrid electric vehicle controllers typically contain a moderate number of parameters (e.g., current battery state-of-charge and torque) that can be tuned using many-objective evolutionary algorithms to solve hybrid electric vehicle controller design problems. The work by Cheng et al.~\cite{Cheng2017} embeds a \emph{preference articulation method} into three evolutionary algorithms (i.e., MOEA/D, NSGA-II, and RVEA) to identify solutions to optimise objectives such as fuel consumption, battery stress for improving the peak performance of the hybrid power unit.
The approach proposed by Rodemann et al.~\cite{Rodemann2015} optimises the fuel consumption of cars with the multi-objective evolutionary algorithm SMS-EMOA~\cite{BEUME20071653}. Drehmer et al.~\cite{Drehmer2015} used particle swarm optimisation and sequential quadratic programming algorithms to optimise the stiffness and damper of the suspension system of cars. Meeruang et al.~\cite{Meeruang} used the Tabu search algorithm to minimise driving time and distance to reduce the fuel consumption of vehicles travelling in cities.

Though using search algorithms for optimisation, these works focus on fuel consumption and vehicle controllers. Instead, \method aims to find minimal changes in vehicle characteristics and increase the chance of putting the vehicle into unsafe situations.

\section{Proposed Approach}\label{sec:approach}
% \IEEEraisesectionheading{\section{Proposed Approach}\label{sec:PROPOSED APPROACH}}

In this section, we first define terminology and present the overall scope in Section~\ref{subsec:definitions}, followed by the safety metrics (Section~\ref{subsec:metrics}). In Section~\ref{subsec:SBApproach}, we introduce \method in detail.

\subsection{Definitions and Scope}\label{subsec:definitions}
In \method, a vehicle under test, i.e., \car, can have a set of ($n$) vehicle characteristics: $\paramsSet = \{C_1, \ldots, C_n\}$. The values of characteristic $C_i$ fall within a given \emph{domain} $D_i = [l_i, u_i]$, with the \emph{range} being $u_i - l_i$. For instance, \car's \mass can range from 2040kg to 2700kg.
%\carp denotes the vehicle with characteristics \params and {\carc} indicates that the characteristics of \car can take concrete values $[v_1, \ldots, v_n]$, {i.e., the original values}.
{\carp denotes the vehicle with characteristics values \params; e.g., \carv indicates that the characteristics of \car take the original values $[v_1, \ldots, v_n]$.}

As defined by Ulbrich et al.~\cite{ScenarioDefined}, a scenario describes the temporal sequence of scene elements, with actions and events of the participating elements occurring within this sequence. Therefore, a scenario \scen characterises the driving environment (e.g., traffic participants, weather, the initial state of \car in terms of initial position) and behaviours of other agents (e.g., NPC vehicles, pedestrians) interacting with \car or static objects such as street signs or traffic cones.

%\subsection{Evaluation metrics}
% explain
When running a scenario (\scen) with a simulator, we can assess the performance of \car regarding safety, comfort, etc. For instance, we can compute the minimum Euclidean distance of \carp to other objects in \scen (e.g., NPC vehicles or pedestrians) via a simulation of \scen to quantify and monitor the safety status of \car. In this paper, we focus on the {\it safety} of \car; however, other aspects, such as comfort and energy consumption, are also worth studying in the future.
% As for comfort and energy consumption, which are subjective and difficult to evaluate , they could be considered as the future work.
% in contrast to comfort and energy consumption, which are subjective and difficult to evaluate
%emergency braking operation

%A suitable metric \safetyDegree proposed in our previous work~\cite{yin} is applied to assess Automatic Emergency Braking (AEB), which is one of the main functions implemented in ADSs. Note that, multifunctional cooperation (e.g., Adaptive Cruise Control, AEB) is required for ADSs to drive safely. Employing the metric \safetyDegree to meassure the scenario we designed can effectively assess vehicle safety status.
%, so the tested ADS can complete this function.

\subsection{Safety Metrics}\label{subsec:metrics}
{There exist various safety metrics in literature as studied by Jahangirova et al.~\cite{jahangirova2021quality}, Mahmud et al.~\cite{mahmud2017application}, etc.} Here, we introduce four of them, any of which can be used to assess the safety during the search of \method. In our experiments, we will instantiate \method with the first one ({\it safety degree}); still, we will also assess the solutions found by \method in terms of the other three metrics.

\textbf{Safety Degree} (\safetyDegree) measures the final distance from \carp to an obstacle (i.e., the NPC vehicle or the pedestrian in our experiments) under the condition that no collision occurs. Otherwise, if a collision occurs, the degree is the negative collision velocity. Formally, we define it as: 
\begin{align}\label{eq:safetyDegree}
\safetyDegree(\res) =
\begin{cases}
minDis(res) & \text{if } minDis(res) > 0\\
$-$ \mathit{collSpeed(res)} & \text{otherwise}
\end{cases}
\end{align}
where \res is the result of a simulation of \scen, i.e., $\res = \simulation(\carp, \scen)$, containing information like the driving path of \carp, its acceleration and velocity over time, and paths followed by NPC vehicles and pedestrians. $\mathtt{minDis}(\res)$ is the minimum distance to the obstacle, and $\mathtt{collSpeed}(\res)$ is the collision velocity. A longer distance of \carp from the obstacle means that \carp is safer. For collision cases, the higher the collision velocity, the more unsafe the situation is. This is measured as the negative value of $\mathtt{collSpeed}(\res)$ because the occurrence of a collision represents a significant loss of control.

Based on the conventional time-to-collision (TTC) metric, \textbf{Time Exposed Time-to-collision (\tet)} and \textbf{Time Integrated Time-to-collision (\tit)} were proposed to assess the safety of \car. They were initially proposed by Minderhoud and Bovy~\cite{MINDERHOUD200189} and later used for ADS and ADAS testing~\cite{TITTETCACC,TETTITDQ}.
TTC is calculated by projecting the relative velocity \velocity on the relative distance \relativeDistance~\cite{Gonner2009ITSC}.
\begin{equation}\label{eq:TTC}
\ttc(t) = \frac{\relativeDistance}{\velocity} 
\end{equation}
%
% where the relative distance \relativeDistance and relative velocity \velocity are calculated as:
% %
% \begin{equation}
% \begin{cases}
% & \relativeDistance = L_{12} - L_{o} \\
% & \velocity = v_{x}\cos\phi + v_{y}\sin\phi \\
% & \tan\phi = \frac{y}{x} \\
% \end{cases}
% \end{equation}
% %
% where $L_{12}$ is the centroid distance between two vehicles (or pedestrians); The included angle with x-axis is $\phi$; $L_{o}$ represents the length of the vehicle.
%For both ADASs and ADSs, large numbers of collisions are not common, so we used the above indicators to measure the impact of variations in vehicle configurations on the exposure to risk in non-collision scenarios during a search process.
%

\tet is defined as the number of time instants $t$, within a specified time interval $H$, in which $\ttc(t)$ is lower than a critical threshold \ttcStar. Thus, the lower a \tet value is, the safer the corresponding situation is considered (over period $H$). To calculate \tet, we assume that TTC at instant \textit{t} remains unchanged for a short time step \timeStep. Hence, there are $T=H$/\timeStep time instants \textit{t} ($t = 0 \ldots T$) to consider: 
\begin{equation}\label{eq:TET}
\begin{split}
& \tetStar = \sum_{t=0}^{T} \delta(t) \cdot \timeStep \\
&\text{with } \delta(t) = \begin{cases}
1 & \text{if } 0 \le \ttc(t) \le \ttcStar \\
0 & \text{otherwise}
\end{cases} \\
\end{split}
\end{equation}
%
%where $\delta(t)$ is a Boolean function.
At instant \textit{t}, if $\ttc(t)$ is between 0 and a specified threshold \ttcStar% (e.g., 1.5s used in our experiment with \carla, see Section~\ref{subsubsec:exDesign}), 
the value of $\delta(t)$ is 1, otherwise 0.

\tit not only considers whether TTC values are below threshold \ttcStar (as TET does), but also \emph{how much}. The higher a \tit value is, the longer \carp is exposed to unsafe TTC values. Thus, in discrete time \timeStep, it is defined as:
\begin{equation}\label{eq:TIT}
\begin{split}
%& \titStar = \sum_{t=0}^{T} \ [\ttcStar-\ttc(t)]\cdot \timeStep \\
%& \forall \ 0 \le \ttc(t) \le \ttcStar
& \titStar = \sum_{\ttc \in \ttcCritical} \ [\ttcStar-\ttc]\cdot \timeStep \\
& \text{with } \ttcCritical =
\left\{\ttc(t) \mid
\begin{array}{l}
t \in \{0, \ldots, T\} \wedge\\
0 \le \ttc(t) \le \ttcStar
\end{array}
\right\}
\end{split}
\end{equation}

{These three metrics are complementary. \tet cumulatively measures how long a potentially dangerous situation lasted in a driving scenario. \tit sums up all the TTC values over a given time duration. They both offer a dynamic and integrated perspective of potential danger over a period of time. Instead, \safetyDegree is a static measure because it does not provide information about the risk progression over time. Hence, the three metrics together offer a comprehensive understanding of ADS safety. }

%\safetyDegree, on the other hand, focuses on the distance between the vehicle and the obstacle when no collision occurs but does not consider the speed of the two objects in the scenario, i.e., One situation is that the vehicle maintains a slow speed and finally stops in front of the pedestrian, another case is that the vehicle at a certain speed and the pedestrian pass each other at close range. Compared to the aboved cases, it is possible to obtain similar distances(i.e., the value of \safetyDegree), but the latter would be more dangerous, because in this case the value of TTC would be lower than the threshold and be recorded in \tet. 

%
\textbf{Average Deceleration} (\aveDece)
Based on deceleration, several metrics have been proposed for assessing the safety of ADSs and ADASs, such as Deceleration Rate to Avoid a Crash (DRAC)~\cite{almqvist1991use}, Crash Potential Index (CPI)\cite{cunto2007microlevel}, and Criticality Index Function (CIF)~\cite{chan2006defining}. {These metrics not only assess potential collision risks based on deceleration rates but also cover safety aspects related to \car's interactions with the environment (e.g., crashes to objects on the road). Since we already cover the safety aspect with the safety-related metrics, we define \textit{aveDece}, which simply measures the average deceleration of \car when changing its \VehicleCharacteristicsSettings, without the need to consider safety-related factors.}

The collection of deceleration starts from the moment \carp receives the braking command from the ADS and stops when the ADS no longer sends out braking commands. Finally, we average the deceleration values during the braking cycle and obtain an \aveDece value with the formula below:
\begin{equation}
\begin{split}
& \mathit{aveDece} = \frac{\sum_{t=0}^{T}\mathit{deceleration}_{t} \cdot \zeta(t)}{\sum_{t=0}^{T} \zeta(t)} \\
& \text{with } \zeta(t) = \begin{cases}
1 & \mathit{Command}_{\mathit{brake}}(t) = \mathit{True} \\
0 & \text{otherwise}
\end{cases} 
\end{split}
\end{equation}
where at instant \textit{t}, $\mathit{deceleration}_{t}$ of \carp can be obtained from the simulator using its APIs. The $\zeta(t)$ value is 1 when the ADS issues the brake command; otherwise, it is 0. In period T, \carp responds to one complete and continuous braking cycle, which involves applying a series of braking commands but no throttle command applied.

\subsection{Optimisation problem and search objectives}\label{subsec:SBApproach}
We aim to find minimum variations \modParamValues to {the original values} \origParamValues of \carv that could decrease its safety to the maximum extent. In other words, we search for situations where, for a given driving scenario \scenSearch, \carv behaves safely, but \carvp does not. We opt to solve this problem with a search-based approach. 

%can be employed to evaluate the impact of configuration variations on the safety. Also, other ADSs (besides Apollo) can be tested. 

% need 
%This section contains two subsections. In subsection 3.2.1, we illustrate how to represent an individual for the search algorithm, and how to derive from it an alternative configuration for \carv. Then, in subsection 3.2.2, we introduce the objective functions that are defined for the search problem. 

\subsubsection{Problem Representation}\label{sec:problemRepresentation}
Search variables are defined as $\searchVars = [x_1, \ldots, x_n]$, where $x_i$ is a value for $C_i \in \paramsSet$ (see Section~\ref{subsec:definitions}). The interval of each variable is given by the domain of the corresponding characteristic, i.e., $x_i \in D_i$. The search variables \searchVars should be assigned values $\modParamValues = [v_1^\prime, \ldots, v_n^\prime]$, which are different from their original values \origParamValues of \carv. That is, we identify with \modParamValues a new \VehicleCharacteristicsSetting different from the original one. 

% changed small
Slight changes to \textit{all} characteristics won't help engineers to perform diagnoses. To be practically meaningful, we employ a \emph{filter}, which defines a threshold for each characteristic to ensure that not all its variations are considered different. Namely, for each characteristic $C_i$, we define a threshold \thresholdI that identifies the minimum variation necessary to be regarded as an assignment $v_i^\prime$ to $C_i$ different from the original one $v_i$. To better define these thresholds, one needs to consult vehicle specifications and rely on domain knowledge. If the absolute difference between $|v_i - v_i^\prime|$ is within the threshold \thresholdI, the characteristic $C_i$ is kept as the original value $v_i$. Otherwise, the characteristic $C_i$ must be considered different. Formally, the filter is defined as follows: 
\begin{equation}\label{eq:changeTh}
\filterV(v_i^\prime) =
\begin{cases}
v_i^\prime & \text{if } |v_i - v_i^\prime|\ > \thresholdI\\
v_i & \text{otherwise}
\end{cases}
\end{equation}

% no change
Therefore, for each individual $\modParamValues = [v_1^\prime, \ldots, v_n^\prime]$ generated by the search, we obtain its filtered version $\filteredModParamValues = [\filterV(v_1^\prime), \ldots, \filterV(v_n^\prime)]$.
Given an individual \modParamValues, we say that characteristic $C_i$ is \emph{selected} and \emph{changed} if the filter keeps the changed value $v_i^\prime$ in the filtered version \filteredModParamValues; otherwise, the characteristic is \emph{not selected} and \emph{not changed}.

%\todo[inline]{I realized that we could have defined this differently, i.e., searching in a given interval and then adding the min or max value (depending on the direction). We should motivate why we do it like this}

\subsubsection{Objective functions}\label{sec:subsubObjective}
%We define three conflicting objectives and specify three corresponding fitness functions that converge to minimize.
%The first objective is to minimise \safetyDegree of a simulation.
The first objective is to minimise the safety of the modified vehicle. As explained in Section~\ref{subsec:metrics}, different safety metrics can be adopted to assess safety; 
%revision{It is a fine-grained safety indicator that can more intuitively reflect the safety distance of vehicle and other traffic participants in the scenario. Compared with \tet and \tit, which pays more attention to unsafe situation, \safetyDegree can reflect the varying degrees of impact of changing \VehicleCharacteristicsSettings on ADS safety, including the situation, in which safety decreases but does not fall below the TTC threshold. Therefore, \safetyDegree is helpful to guide the entire convergence process of the search algorithm.} 
In the experiment of this work, we use the \safetyDegree (see Eq.~\ref{eq:safetyDegree}).
So, given the individual \filteredModParamValues, the fitness function is defined as:
\begin{equation}\label{eq:fitnessSafety}
\resizebox{.95\hsize}{!}{\fitSaf(\filteredModParamValues) = \safetyDegree(\res) \quad \text{with }\res = \simulation(\carvpp, \scenSearch)}
\end{equation}
where \carvpp is \carp configured with a set of filtered characteristic values \filteredModParamValues and \scenSearch is a given driving scenario. The fitness evaluation requires the simulation of \carvpp in \scenSearch running in a simulator.
%, and the simulation time cannot be negligible. Therefore, the fitness computation is the main factor affecting our approach's scalability.
%In our pilot experiment, we observed that results in terms of \safetyDegree are consistent with those of \tet and \tit in vehicle safety assessment. Hence, we only selected \safetyDegree as the safety measure for this objective. This observation is also valid for the formal empirical study, as reported in Section~\ref{subsubsec:RQ2}.

% no
To find a not-too-different \VehicleCharacteristicsSetting, we define the second objective that minimises the maximum percentage variation between characteristic values \origParamValues and the filtered characteristic values \filteredModParamValues. The fitness function is therefore defined as:
\begin{equation}\label{eq:valueDiff}
\fitDiff(\filteredModParamValues) = \max_{i \in \{1, \ldots, n\}} \frac{|v_i - v_i^{\prime\prime}|}{v_i}
\end{equation}

To further restrict vehicle characteristics variations, we pay attention to the number of changed characteristics. Therefore, we specify the third objective as minimising the number of changed characteristics. Namely, the corresponding fitness function is defined as:
\begin{equation}\label{eq:fitNumDiff}
\fitNumDiff(\filteredModParamValues) = |\{i \in \{1, \ldots, n\} \mid v_i^{\prime\prime} \neq v_i\}
\end{equation}

The rationale behind these two last objectives is that, in practice, variations in \VehicleCharacteristicsSettings are mainly due to production errors, wear and tear, usage contexts (e.g., overload), etc. Such variations are not large, and the number of \VehicleCharacteristicsSettings involved in a specific application context is often small. Furthermore, from an engineer's perspective, it is more helpful to have a small number of \VehicleCharacteristicsSettings changed concurrently, as it will simplify the diagnostic analyses to some extent. 

\section{Empirical Evaluation}\label{sec:evaluation}
% \IEEEraisesectionheading{\section{EMPIRICAL EVALUATION}\label{sec:EMPIRICAL EVALUATION}}
%To evaluate \method, we conduct and present an empirical evaluation. Three parts are divided in this section. 
Section~\ref{subsec:exDesignAndExecution} presents the design and execution of our experiments, followed by the research questions in Section~\ref{subsec:researchQuestion}, statistical tests used for answering RQs in Section~\ref{subsec:StatisTestsandMetric} and experiment results and analyses in Section~\ref{subsec:researchResult}.

\subsection{Experiment Design and Execution}\label{subsec:exDesignAndExecution}

\subsubsection{Experiment Design}\label{subsubsec:exDesign}

{\it \textbf{Simulators.}} We employ two autonomous driving simulators: \carla(version 0.9.10) and \lgsvl (version 2021.1). \carla, as an open-source simulator, is a plugin to Unreal Engine\footnote{https://www.unrealengine.com/} -- an open-source video game engine. \carla leverages this engine to simulate vehicles' physics and generate simulated sensor data from cameras, LiDAR, etc. \carla is capable of simulating behaviours of various agents such as NPC vehicles, pedestrians, and motorcyclists. It is also equipped with a set of APIs to support the development and testing of ADSs. %simulating virtual vehicles and driving environments. 
\lgsvl\footnote{{Though the active development of \lgsvl has been suspended, there exist modified versions of \lgsvl in the community, which can be used by \method. Moreover, \method is independent of simulators; consequently, it can be tailored for other available simulators.}} is also open source and was developed with Unity\footnote{https://unity.com/}. \lgsvl simulates a driving environment (e.g., traffic and physical environment), sensors involved (e.g., camera, LiDAR, GPS, and Radar), and vehicle dynamics. 
%\lgsvl proposes an out-of-the-box solution for testing ADSs. And it is integrated into a number of platforms, making the entire system easy to test and verify. 
Notably, \lgsvl supports a variety of middleware, which provides communication and resource management services for Autopilot software (e.g., ROS1, ROS2, and Cyber RT messages, which help connect itself to Baidu Apollo), a well-known autonomous driving stack.

{\it \textbf{Subject Systems.}}
One of the two employed ADSs is World On Rails (WOR)~\cite{WOR}, an end-to-end ADS implemented in \carla based on model-based RL. %In contrast to other model-based reinforcement learning methods, this work builds a forward world model composed of the controllable ego-agent (i.e., a component that reacts to the agent’s commands) and a passively moving environment to simplify policy learning. 
WOR can handle the case of end-to-end urban driving, including lane keeping, traffic light detection, pedestrians crossing roads, and obstacle avoidance. WOR has been evaluated on various benchmarks (i.e., CARLA public leaderboard\footnote{https://leaderboard.carla.org/leaderboard/}, the Town05 benchmark~\cite{TransFuser}, NoCrash benchmark~\cite{NoCrash}, and CARLA 42 Routes benchmark~\cite{42RoutesBenchmark}). Compared with other state-of-the-art end-to-end ADSs (e.g., LBC~\cite{chen2019lbc}, TransFuser~\cite{TransFuser}), WOR has achieved better performance~\cite{Shao2022SafetyEnhancedAD}.
We adopted Apollo as the second ADS under test and deployed it on \lgsvl. According to the six levels of autonomy in a vehicle defined by the Society of Automotive Engineers (SAE)~\cite{sae2014definitions}, Apollo is an industrial-level ADS, which can reach the L4 level~\cite{zhong2021survey}. With multiple deep learning models, Apollo can handle data from cameras, Lidar, Radar, IMU, GPS, and high-resolution maps. %(e.g., a new prediction model to handle the changing conditions of the complex road and junction scenarios~\cite{}), and 
%Apollo can also handle complex end-to-end urban driving tasks.

%, moving closer to fully autonomous urban road driving.

{\it \textbf{Autonomous Vehicles Under Test.}} With \carla, the virtual vehicle is LincolnMkz2017, as WOR was trained with it. With \lgsvl, we selected Lincoln2017MKZ as the virtual vehicle on which Apollo was deployed. We used Lincoln2017MKZ because it is commonly used in ADS research, such as the works by Xu et al.~\cite{LincolnR&D} and Vegamoor et al.~\cite{LincolnVegamoor2022StringSO}. Both virtual vehicles employed in the two simulators correspond to the real-world 2017 Lincoln MKZ vehicle but implement different vehicle dynamics models.

{\it \textbf{Driving Scenarios.}}
To test the two ADSs, we designed one driving scenario for each simulator. These two scenarios require the ADSs to take action to avoid unsafe situations such as collisions. 
%According to the NHTSA 37 pre-Crash Scenarios~\cite{Najm2007PreCrashST}, two common driving scenarios are defined in \carla and \lgsvl, respectively.
As described by Najm et al.~\cite{najm2007pre}, the NHTSA 37 pre-crash scenarios provide references for researchers to determine which traffic safety scenarios should be considered critical. We selected two of them. %, i.e., \textit{Lead Vehicle Stopped} and \textit{Pedestrian Crash With Prior Vehicle Manoeuvre}, and implemented them in \lgsvl and \carla, respectively.

For \carla, we selected the \textit{Pedestrian Crash With Prior Vehicle Manoeuvre} scenario, as pedestrians are road users most likely to be injured or even killed in traffic collisions~\cite{world2018global}.
%Therefore, we experiment with driving scenarios involving pedestrians with \carla.
Particularly, we selected the virtual driving environment $\mathit{Town01}$ with the following driving scenario: \textit{LincolnMkz2017 + WOR} drives from a specified starting point to an ending point (destination), and a pedestrian is spawned at a fixed position in the vehicle's driving path and follows a predetermined trajectory to cross the road. When facing the pedestrian crossing the road, \textit{LincolnMkz2017 + WOR} is expected to make an emergency stop to avoid hitting the pedestrian in time and wait until the pedestrian crosses the road before continuing the driving to the destination.
For \lgsvl, instead, we selected the \textit{Lead Vehicle Stopped} scenario.\footnote{As observed in the experiment reported by Lu et al.~\cite{lu2022learning} and our pilot experiment, Apollo 5.0 in \lgsvl exhibited unsafe behaviour in scenarios of pedestrians crossing roads even with \OriVehicleCharacteristicsSetting. Hence, we could not use the same scenario (involving pedestrians) as for \carla.} Namely, we designed a scenario involving NPC vehicles: \textit{Lincoln2017MKZ + Apollo} starts from a specific location on the map and drives on a two-lane highway until it encounters two non-ego vehicles parked on the driving path ahead, then makes an emergency stop or hits them. As the virtual environment, we used the $BorregasAve$ map of a real-world suburban street block in Sunnyvale, CA.

With \carla, we designed two weather conditions: \precSetCarlaSun and \precSetCarlaRain. {This is because WOR is a vision-based ADS. Rainy weather affects the camera image and hence challenges the ADS. Rain also affects the friction between tires and roads, hence vehicle behaviour. }
With \lgsvl, we only experimented with the sunny weather condition, named \precSetLgsvl, because we observed no noticeable impact of weather conditions on the safety of Apollo when running it in \lgsvl in our pilot experiment. %Also, it was already highly costly for us to conduct experiments with the current design. 

%Note that, we designed different scenarios in the two simulators. It is not our intention to compare the two simulators, for the reason that both the selected virtual vehicles and their configurations in the two simulators are not identical. Testing different ADSs with different simulators and scenarios is to extend the generality of our proposed method.
{Mahmud et al.~\cite{mahmud2017application} surveyed settings of the TTC threshold \ttcStar in various situations where traffic conflicts occur frequently. For example, the study by Host~\cite{van1991TTC1_5} shows that when a vehicle approaches an intersection, the desirable TTC threshold \ttcStar is 1.5s. In \precSetCarlaSun and \precSetCarlaRain, pedestrians crossing the road form a cross-interaction situation while the vehicle goes straight ahead. Hence, we set \ttcStar 1.5s based on the result from Host~\cite{van1991TTC1_5}. 
According to various studies~\cite{farah2008modelTTC3,aashto2001policyTTC3,hegeman2008assistedTTC3}, the recommended TTC threshold for traffic conditions on 2-lane rural roads is 3s. Our scenario in \precSetLgsvl is similar to this. Hence, we set the TTC threshold of 3s for the scenario in \precSetLgsvl.}
%Considering that the scenario of a pedestrian crossing the road in \carla is more dangerous, we choose 1.5s as TTC threshold \ttcStar when calculating \tet and \tit. For the lower risk scenario \precSetLgsvl, we choose the TTC threshold of 3s. These TTC thresholds have been studied in various contexts~\cite{van1991TTC1_5, farah2008modelTTC3,aashto2001policyTTC3,hegeman2008assistedTTC3}. 

{{\it \textbf{Vehicle Characteristics Settings.}}}
Cheng et al.~\cite{Cheng2017}, Minder et al.~\cite{Minder2022} and Drehmer et al.~\cite{Drehmer2015} reported that torque, mass, and radius are common vehicle characteristics being studied. Therefore, for the combination of \textit{LincolnMkz2017} and \carla, we consider them, in addition to the nine characteristics commonly employed to simulate vehicle dynamics in the \carla simulator (Table~\ref{tab:paramsVehicleInCarla}). 
\begin{table}[!tb]
\caption{Characteristics of the vehicle}
\begin{subtable}{1\columnwidth}
\centering
\caption{Virtual vehicle in \carla}
\label{tab:paramsVehicleInCarla}
\resizebox{\columnwidth}{!}{
\begin{tabular}{ccc}
\toprule
Characteristic $C_i$ & Original Value \origParamValues & Domain $D_i = [l_i, u_i]$\\
\midrule
\maxRpm(r/min) & 5800 & {[}4200, 7000{]} \\
\dampingRateFullThrottle(kg*{m$^2$}/s) & 0.15 & {[}0.1, 0.2{]} \\
\dampingRateZeroThrottleClutchEngaged(kg*{m$^2$}/s) & 2 & {[}1.0, 3.0{]} \\
\dampingRateZeroThrottleClutchDisengaged(kg*{m$^2$}/s) & 0.35 & {[}0.2, 0.4{]}\\
\gearSwitchTime(s) & 0.5 & {[}0.3, 0.6{]} \\
\clutchStrength(kg*{m$^2$}/s) & 10 & {[}8.0, 12.0{]} \\
\mass(kg) & 2404 & {[}2040, 2700{]} \\
\dragCoefficient(-) & 0.3 & {[}0.2, 0.5{]} \\
\tireFriction(-) & 3.5 & {[}1.0, 3.9{]} \\
\dampingRate(kg*{m$^2$}/s) & 0.25 & {[}0.20, 0.30{]} \\
\radius(cm) & 35.5& {[}31.7, 37.0{]} \\
\maxBrakeTorque(N*m) & 1500 & {[}1200, 1650{]} \\
\bottomrule
\end{tabular}
}
\begin{flushleft} 
{\footnotesize *\maxRpm: maximum RPM of the vehicle's engine - \dampingRateFullThrottle: damping ratio when the throttle is maximum - \dampingRateZeroThrottleClutchEngaged: damping ratio when the throttle is zero with the clutch engaged - \dampingRateZeroThrottleClutchDisengaged: damping ratio when the throttle is zero with the clutch disengaged - \gearSwitchTime: switching time between gears - \clutchStrength: clutch strength of the vehicle - \mass: mass of the vehicle - \dragCoefficient: drag coefficient of the vehicle's chassis - \tireFriction: scalar value that indicates the friction of the wheel - \dampingRate: damping rate of the wheel - \radius: radius of the wheel - \maxBrakeTorque: maximum brake torque.}
\end{flushleft}
\end{subtable}

\vspace{10pt}

\begin{subtable}{1\columnwidth}
\centering
\caption{Virtual vehicle in \lgsvl}
\label{tab:paramsVehicleInLgsvl}
\resizebox{\columnwidth}{!}{
\begin{tabular}{ccc}
\toprule
Characteristic $C_i$ & Original Value \origParamValues & Domain $D_i = [l_i, u_i]$\\
\midrule
\mass(kg) & 2120 & {[}2000, 2500{]} \\
\wheelMass(kg) & 30 & {[}20, 60{]} \\
\radius(m) & 0.35& {[}0.30, 0.39{]} \\
\maxRpm(r/min) & 8299 & {[}6000, 13000{]} \\
\minRpm(r/min) & 800 & {[}600, 1100{]} \\
\maxBrakeTorque(N*m) & 3000 & {[}2500, 3150{]} \\
\maxMotorTorque(N*m) & 450 & {[}400, 550{]} \\
\maxSteeringAngle(N*m) & 39.4 & {[}30, 50{]} \\
\tireDragCoeff(-) & 4 & {[}2, 5{]} \\
\wheelDamping(kg*{m$^2$}/s) & 1 & {[}0.15, 1.50{]} \\
\shiftTime(s) & 0.4 & {[}0.2, 0.6{]} \\
\tractionControlSlipLimit(s) & 0.8 & {[}0.65, 0.95{]} \\
\bottomrule
\end{tabular}
}
\end{subtable}
\begin{flushleft} 
{\footnotesize *\mass: mass of the vehicle - \wheelMass: mass of the wheel - \radius: radius of the wheel - \maxRpm: maximum RPM of the vehicle's engine - \minRpm: minimum RPM of the vehicle's engine - \maxBrakeTorque: maximum brake torque - \maxMotorTorque: maximum Motor torque - \maxSteeringAngle: maximum steering angle - \tireDragCoeff: tire resistance coefficient - \wheelDamping: damping rate of the wheel - \shiftTime: time interpolation of gear shifts - \tractionControlSlipLimit: traction control limits torque based on wheel slip - traction reduced by amount when slip exceeds the tractionControlSlipLimit.}
\end{flushleft}

\end{table}
\carla provides APIs to configure these 12 characteristics. Table~\ref{tab:paramsVehicleInCarla} presents the original characteristic values of \textit{LincolnMkz2017}, along with the ranges of the characteristics. Some of these ranges are from~\cite{NVIDIA}.
%In our experiments, any characteristic's value within its range is considered valid.
For \textit{LincolnMkz2017} with the \lgsvl simulator, we also selected 12 characteristics as shown in Table~\ref{tab:paramsVehicleInLgsvl}, {among which $\mathit{max\_rpm}$, $\mathit{mass}$, and $\mathit{radius}$ are the same as for \carla}. Since \lgsvl does not provide APIs to change values of the vehicle characteristics, based on the guidelines~\cite{rong2020lgsvl}, we developed our own APIs to make this possible. %{Also, in this case, we checked that the minimum and maximum variations of the characteristics do not lead to abnormal behaviours of the vehicle.}
{The two simulators implement different vehicle dynamic models, i.e., \carla uses the embedded Nvidia-phyX, and \lgsvl is based on Unity's self-made model import. As a result, there are differences in the studied characteristics. %Out of the 12 characteristics, three characteristics, i.e., $max\_rpm$, $mass$, and $radius$ are the same. %Note that we do not intend to directly compare the two simulators in this paper. Instead, we study the impact of variations of \VehicleCharacteristicsSettings on the safety of ADS running on each simulator.
}
{In addition, as shown in the official documents and source code of \carla~\cite{CARLA} and \lgsvl~\cite{rong2020lgsvl}, these vehicle characteristic settings affect the calculation of the vehicle dynamics model and hence the behaviour of the virtual vehicle, which in turn affects the ADS's control over it. For example, \tireFriction sets the coefficient of friction between the vehicle tires and the road.} 

{In Tables~\ref{tab:paramsVehicleInCarla} and~\ref{tab:paramsVehicleInLgsvl}, domain $D_i$ specifies the minimum and maximum values, i.e., $l_i$ and $u_i$ for each characteristic. For both simulators, to guarantee that the applied changes do not lead to abnormal vehicle behaviours (e.g., the vehicle never moving nor breaking), we simulated the scenarios using the maximum and minimum values of each characteristic, and we checked that the vehicle progresses in the scenario (with various levels of safety). Hence, we consider all values of the characteristics in their ranges valid.}

{Though some of the characteristics are correlated, the natural evolution of these configuration parameters is less correlated. For instance, heavier vehicles demand more braking force. However, the maximum brake torque of a vehicle's braking system defines the maximum force the brake can apply to stop the vehicle. The change of the intended initial values of \textit{mass} and \textit{maxBrakeTorque} due to production errors and wear and tear are, however, independent. For example, the increase in mass might be caused by adding passengers or heavy equipment, while the braking system's effectiveness can degrade over time due to wear and tear.}

{{\it \textbf{Filter Implementation.}}}
As explained in Section~\ref{sec:problemRepresentation}, to restrict vehicle characteristic variations, we implemented a filter that takes all characteristic values as input, calculates the percentage value change of each characteristic (when compared with its original value), and then compares it with a predefined threshold. If the amount of the change is greater than threshold \thresholdI, the \VehicleCharacteristicsSetting is updated with the changed value; otherwise, \OriVehicleCharacteristicsSetting is kept. {The filter is needed to prevent the search from generating many small changes that do not impact the vehicle's safety much.}

{However, defining threshold \thresholdI for each characteristic within its domain $D_i$ remains an open issue. As discussed in Section~\ref{sec:problemRepresentation}, one possible solution is to look into available vehicle specifications and other domain knowledge, which, unfortunately, we do not have. Therefore, we looked into relevant literature. Yin et al.~\cite{yin} proposed a preliminary solution: defining characteristic $C_i$'s \thresholdI as percentage variation \precision (named as \emph{precision}) of the range of the characteristic domain $D_i$ (see Eq.~\ref{eq:thresholdIeq}).
{
\begin{align}\label{eq:thresholdIeq}
\thresholdI = \precision \times ( u_i - l_i ) \quad \text{with } D_i = [l_i, u_i]
\end{align}}
Yin et al.~\cite{yin} experimented with two sets of precision values and observed that the results with a greater precision value led to a more severe impact on the vehicle's safety. Based on these results, we initialise \precision, in this study, with the greater values (as shown in Eq.~\ref{eq:domain}).}

\begin{align}\label{eq:domain}
\precision =
\begin{cases}
0.01 & D_i \in {[}1000,+\infty{)}\\
0.02 & D_i \in {[}100,1000{)}\\
0.04 & D_i \in {[}1,100{)}\\
0.08 & D_i \in {[}0,1{)}
\end{cases}
\end{align}
For instance, for \mass, {the \precision is 0.02 because its $D_i$ falls into the range of 100 to 1000.} Consequently, $\mathit{Th}_{\mass}$ is $0.02 \times (2700-2040) = 13.2$ for \precSetCarlaSun.

{\it \textbf{Settings of the Search Algorithms.}}
For search algorithm implementation, we employed the jMetalPy framework~\cite{benitez2019jmetalpy} and used its default implementation of NSGA-II~\cite{deb2002fast}. The main reason for selecting NSGA-II is that it is a widely applied multi-objective search algorithm in search-based software engineering and fits our context. The default setting of NSGA-II in jMetalPy is that it has the crossover rate of the Simulated binary crossover (SBX) operation as 0.9, and the mutation rate of the polynomial mutation operation (1/12) being equal to the reciprocal of the number of variables. Since running simulations with the two simulators is time-consuming, we conducted a pilot study to understand the convergence of NSGA-II and concluded that it roughly converges at the 100th generation. Based on the pilot study results, we set the number of generations to 100 for the experiments with both \carla and \lgsvl. Regarding the population size, we set it to 50 for the experiments with \carla and 30 for the experiments with \lgsvl. The termination criterion is the number of fitness evaluations being 5000 (3000) (i.e., 50 × 100, 30 × 100) for all the experiments. 
%Because the experimental cost is expensive, we suggest different cases (i.e., vehicle dynamics models and ADS on different platforms) could have different convergence trends of generation and population size. 

{{\it \textbf{Baselines.}}}
{We chose the random search algorithm (\textit{RS}) implemented by jMetalPy as a baseline to verify whether the search is needed. For a fair comparison, RS uses the same number of fitness evaluations as its termination condition.}

{Since different ADS testing works rely on fuzzing, we also implemented a mutation-based fuzzer, named \safeFuzzer, as the second baseline. Its pseudo-code is reported in Algorithm~\ref{alg:algo_fuzzer}.}
\begin{algorithm}[!tb]
\SetKwInOut{Input}{input}\SetKwInOut{Output}{output} 
\footnotesize
\Input{$\mathit{pop\_size}$: population size\; $\mathit{range_{next}}$: upper and lower bounds of each vehicle characteristic\; $\mathit{maxNum_{next}}$: maximum number of characteristics to change in the current 5 generations\; 
} 
\Output{$\mathit{Optimal_{sol}}$}
\BlankLine 
$\mathit{current\_eva}\leftarrow 0$\; 
$\mathit{converged} \leftarrow \mathit{False}$\;
$\mathit{all}_{\modParamValues}\leftarrow [];$\tcp*[f]{\scriptsize All \VehicleCharacteristicsSettings and simulation results}

$\mathit{curFive}_{\modParamValues} \leftarrow [];$\tcp*[f]{\scriptsize \VehicleCharacteristicsSettings of the last five generations}

\While(\tcp*[f]{\scriptsize Termination conditions}){$\mathit{current\_eva} \le \mathit{max\_eva}$ and $\mathit{converged} == \mathit{False}$}{
$\mathit{VCSs}\leftarrow \mathit{MutationGen}(\mathit{range_{next}}, \mathit{maxNum_{next}})$\label{line:MutationGen}

\For(\tcp*[f]{\scriptsize Each generation}){each set of \modParamValues in VCSs}{
$\safetyDegree, \tet, \tit, \aveDece\leftarrow \mathit{Simulation}(\modParamValues)$

$\mathit{all}_{\modParamValues}.\mathit{append}(\modParamValues,\safetyDegree)$

$\mathit{curFive}_{\modParamValues}.\mathit{append}(\modParamValues)$

$\mathit{current\_eva}\leftarrow \mathit{current\_eva}+1$
}
\If{$\mathit{current\_eva} > 500$}{
$\mathit{Normalization}(\mathit{all}_{\modParamValues})$\label{line:normalization}

$\mathit{fitnessScore}\leftarrow \mathit{CalFitnessS}(\mathit{all}_{\modParamValues})$\label{line:fitness}

$\mathit{converged}\leftarrow \mathit{IsConverge}(\mathit{all}_{\modParamValues},\mathit{fitnessScore})$
}
\If{$\mathit{current\_eva}\ \%\ (\mathit{pop\_size}*5) == 0$}{
$\mathit{range_{next}}, \mathit{maxNum_{next}} \leftarrow \mathit{Update}(\mathit{curFive}_{\modParamValues}, \mathit{range}_{\mathit{next}}, \mathit{maxNum}_{\mathit{next}})$\label{line:updateRangeNext} 

$\mathit{curFive}_{\modParamValues}\leftarrow []$
}

}
$\mathit{Normalization}(\mathit{all}_{\modParamValues})$

$\mathit{fitnessScore}\leftarrow \mathit{CalFitnessS}(\mathit{all}_{\modParamValues}) $

$\mathit{Opimital_{sol}}\leftarrow \mathit{FindSolution}(\mathit{all}_{\modParamValues}, \mathit{fitnessScore})$

\caption{{The \safeFuzzer baseline}}
\label{alg:algo_fuzzer} 
\end{algorithm}
{Overall, \safeFuzzer mutates existing \VehicleCharacteristicsSettings to generate new ones, guided by the three objectives (Section~\ref{sec:subsubObjective}).
The population size ($\mathit{pop\_size}$) of \safeFuzzer is 50, the same as for NSGA-II. Initially, the value of each vehicle characteristic (constrained by its ranges (Tables~\ref{tab:paramsVehicleInCarla} and \ref{tab:paramsVehicleInLgsvl})) is randomly generated. Then, during each generation, it randomly selects the number of characteristics to change and decides which vehicle characteristics to mutate and how much to change.}

{
Every five generations, \safeFuzzer analyzes the generated \VehicleCharacteristicsSettings that reduce safety compared to \OriVehicleCharacteristicsSetting, i.e., \fuzzerTestCase. Then, it identifies cases where the value of a selected characteristic $C_i$ is changed, either getting larger or smaller in \fuzzerTestCase, and calculates the average of values that are getting larger and the average of those getting smaller, and then sets these two averages as the next iteration's upper and lower bounds of the domain (update of $\mathit{range_{next}}$ at Line~\ref{line:updateRangeNext}). If no change occurs for a characteristic, its original domain is kept. These steps aim to achieve the second objective \fitDiff. 
\safeFuzzer sets the average number of changed characteristics in \fuzzerTestCase as the maximum number of characteristics (update of $\mathit{maxNum_{next}}$ at Line~\ref{line:updateRangeNext}) that can be adjusted for the next five generations so that the algorithm converges towards selecting a smaller number of characteristics to change, aiming to achieve the third objective \fitNumDiff. Then values of $\mathit{range_{next}}$ and $\mathit{maxNum_{next}}$ are taken by the function $\mathit{MutationGen}()$ to generate new \VehicleCharacteristicsSettings (Line~\ref{line:MutationGen}).}

{\safeFuzzer employs a linear weighting method. After each generation, the value of each objective is normalised and weighted with pre-defined weights (Line~\ref{line:normalization}) to calculate the final \fitnessScore (Line~\ref{line:fitness}). In this paper, the weights of $\fitSaf(\filteredModParamValues)$, $\fitDiff(\filteredModParamValues)$ and $\fitNumDiff(\filteredModParamValues)$ are set 6:1:1. Setting \fitSaf's weight 6 is for encouraging \safeFuzzer to generate \VehicleCharacteristicsSettings that lead to unsafe situations. 
\safeFuzzer considers the top 50 \VehicleCharacteristicsSettings with the highest fitness scores, i.e., the optimal solutions. Regarding the termination criterion, we adopt the one applied in AV-FUZZER~\cite{LiISSRE20}: %the current generation does not increase anymore compared with a running average of the last five generations. Thus, we define it as: 
it stops if the optimal solutions are not updated for five consecutive generations.
}

\subsubsection{Experiment Execution}
%Due to the uncertainty of ADSs, one execution of the scenario is non-deterministic, which is to be expected, while the environment configuration in the scenario behaves in the same way in repeated executions. 
Before the experiments, the selected two virtual vehicles with their \OriVehicleCharacteristicsSetting from the two simulators were run in each scenario to ensure that no unsafe behaviour was observed, indicating that with \OriVehicleCharacteristicsSetting, the vehicles behaved as expected. During the experiments, each algorithm (i.e., NSGA-II, RS {and \safeFuzzer}) was run 30 times, as suggested by Arcuri et al.~\cite{Arcuri2011}. %and generate \VehicleCharacteristicsSettings which were applied to configure the virtual vehicles in the two simulators. In the end, we compute \safetyDegree, \tet, \tit, and \aveDece at the end of each time scenario execution.
For each individual generated by the search, we initialise the simulator with the identified characteristics \VehicleCharacteristicsSettings and simulate the scenario; at the end of the scenario execution, we compute the adopted safety metric \safetyDegree that is used in the fitness function \fitSaf (see Eq.~\ref{eq:fitnessSafety}).\footnote{In practice, we also compute \tet, \tit, and \aveDece, so that we can assess them in the evaluation.}

The experiments have been executed on a Linux machine, 2.2 GHz Intel Xeon CPU, and 150GB of RAM.

\subsubsection{Data Availability}\label{subsubsec:DataAvailability}
The replication package is available in the GitHub repository: \url{https://github.com/simplexity-lab/SAFEVAR}. %The following contents are provided in the replication package: 1) The scenario code designed in \carla and the code combining the extended WOR and NSGA-II algorithm. 2) The \lgsvl simulator with our own APIs deployed and the code of scenario designed in \lgsvl combined with the NSGA-II algorithm. 3) The dataset containing all the raw data for the analyses (Section~\ref{subsec:researchResult}).

\subsection{Research Questions}\label{subsec:researchQuestion}
We identified the following Research Questions (RQs):
\begin{compactitem}
\item {{\bf RQ1}: How effective is \method in generating \VehicleCharacteristicsSettings with a higher chance of putting the vehicle into unsafe situations than \OriVehicleCharacteristicsSetting? With this RQ, we want to know whether it makes sense to vary \VehicleCharacteristicsSettings.} 
%With this RQ, we want to study the effectiveness of \method in finding safety-critical \VehicleCharacteristicsSettings. 
\item {{\bf RQ2}}: 
% Does NSGA-II perform significantly better than RS? This RQ aims to test if there is a need for the search.
{To what extent can \method perform significantly better when using NSGA-II than RS or \safeFuzzer in generating \VehicleCharacteristicsSettings? This RQ tests how much NSGA-II outperforms the baselines.}
\item {\bf RQ3}: Which combinations of the vehicle characteristics have a higher chance of putting the vehicle into unsafe situations, and what are the variations of their values? This RQ helps identify critical characteristics and their interactions and is split into three sub-RQs.
\begin{compactitem}
\item {\bf RQ3.1}: How is the overall distribution of the numbers of selected characteristics (out of the 12 vehicle characteristics)? This RQ helps to gain an initial understanding of how effective \method is in minimising the number of characteristics to change.
\item {\bf RQ3.2}: What characteristics are top-ranked regarding times being selected by the search? This RQ helps to identify characteristics frequently selected by the search in various solutions (e.g., 3-characteristic-selected solutions, 5-characteristic-selected solutions).
\item {\bf RQ3.3}: What are value variations of characteristics selected by the search? This RQ aims to discover how much changes to \VehicleCharacteristicsSettings will lead to the degradation of vehicle safety.
\end{compactitem}
% \item {\bf RQ3}: Under different weather conditions in simulations, will the selection of critical characteristics and the scope of characteristics variations be different? (This RQ aims to discover whether even minor changes of the selected parameters in extreme weather can lead to more serious consequences.)
% \item {\bf RQ4}: What are the combinations of configurable parameters that are top ranked in terms of times being selected by the search? This RQ aims to find the possible combinations of parameters whose variations have a stacking effect.
\end{compactitem}

\subsection{Statistical Tests and Metrics}\label{subsec:StatisTestsandMetric}

%As shown in Table~\ref{tab:StaTestsAndMetrics}, we report the statistical tests needed in RQs.

{To answer RQ2, as suggested by Arcuri et al.~\cite{Arcuri2011}, we first perform the Mann-Whitney U test to test if there exists a significant difference between NSGA-II and RS (or \safeFuzzer) regarding IGD and each safety metric (i.e., \safetyDegree, \tet, \tit and \aveDece, see Section~\ref{subsec:definitions}), with the significance level of 0.01. If two algorithms are judged to be significantly different, we calculate the Vargha and Delaney effect size \Atwelve. $\Atwelve>0.5$ indicates that the value of the metric produced by NSGA-II is likely higher than the one produced by RS (or \safeFuzzer); otherwise, RS (or \safeFuzzer) likely produces a higher value. 
}
{As proposed by Ali et al.~\cite{ALIQI}, we employ IGD to measure the performance of NSGA-II, RS, and \safeFuzzer. Specifically, IGD computes the distance of the Pareto fronts generated by the algorithms from the reference Pareto front with a smaller IGD value, indicating a better performance. For each setting, a reference \paretoFront is computed by merging the Pareto fronts of all the 30 runs of RS, \safeFuzzer, and NSGA-II. Then, we compare the 30 IGD values of RS, \safeFuzzer, and NSGA-II.}

Similarly, for RQ1, we used the Mann-Whitney U test and the Vargha and Delaney effect size to compare simulation results with \OriVehicleCharacteristicsSetting with those with \VehicleCharacteristicsSettings generated by NSGA-II{, RS or \safeFuzzer} in terms of all the safety metrics.

\subsection{Results and Analyses}\label{subsec:researchResult}

\subsubsection{Results for RQ1}\label{subsubsec:RQ2}
\begin{table*}[!tb]
\centering 
\caption{Result of comparing \OriVehicleCharacteristicsSetting and \VehicleCharacteristicsSettings generated by NSGA-II{, RS and \safeFuzzer} -- RQ1}
\label{tab:OrigVSNSGAVSRSVSFUZZER}

\renewcommand{\arraystretch}{1.5}
\begin{tabular}{ccc|c|c|c|c}
\hline
\multicolumn{1}{c}{\multirow{2}[0]{*}{Case}} & \multicolumn{2}{c}{\multirow{2}[0]{*}{Algorithm}} 
& \safetyDegree & \tet & \tit & \aveDece \\
\cline{4-7}
& \multicolumn{2}{c}{} & p-value/{1-\Atwelve} & p-value/\Atwelve & p-value/\Atwelve & p-value/{1-\Atwelve} \\

\hline
\multicolumn{1}{c}{\multirow{3}[0]{*}{\precSetCarlaSun}} & \multicolumn{1}{c}{\multirow{3}[0]{*}{\OriVehicleCharacteristicsSetting}} & NSGA-II & \textbf{$<$.01}/{\textbf{0.067}} & \textbf{$<$.01}/\textbf{0.067} & \textbf{$<$.01}/\textbf{0.123} & \textbf{$<$.01}/{\textbf{0.017}} \\

& \multicolumn{1}{c}{} & RS & 
\textbf{$<$.01}/{\textbf{0.170}} & \textbf{$<$.01}/\textbf{0.155} & \textbf{$<$.01}/\textbf{0.222} & \textbf{$<$.01}/{\textbf{0.041}} \\

& \multicolumn{1}{c}{} & \safeFuzzer & 
\textbf{$<$.01}/{\textbf{0.277}} & \textbf{$<$.01}/\textbf{0.154} & \textbf{$<$.01}/0.435 & \textbf{$<$.01}/{\textbf{0.004}} \\

\cline{1-7}
\multicolumn{1}{c}{\multirow{3}[0]{*}{\precSetCarlaRain}} & \multicolumn{1}{c}{\multirow{3}[0]{*}{\OriVehicleCharacteristicsSetting}} & NSGA-II & \textbf{$<$.01}/{\textbf{0.005}} & \textbf{$<$.01}/\textbf{0.018} & \textbf{$<$.01}/\textbf{0.011} & \textbf{$<$.01}/{\textbf{0.020}} \\

& \multicolumn{1}{c}{} & RS & 
\textbf{$<$.01}/{\textbf{0.057}} & \textbf{$<$.01}/\textbf{0.073} & \textbf{$<$.01}/\textbf{0.092} & \textbf{$<$.01}/{\textbf{0.052}} \\

& \multicolumn{1}{c}{} & \safeFuzzer & 
\textbf{$<$.01}/{\textbf{0.000}} & \textbf{$<$.01}/\textbf{0.000} & \textbf{$<$.01}/\textbf{0.000} & \textbf{$<$.01}/{\textbf{0.106}} \\

\cline{1-7}
\multicolumn{1}{c}{\multirow{3}[0]{*}{\precSetLgsvl}} & \multicolumn{1}{c}{\multirow{3}[0]{*}{\OriVehicleCharacteristicsSetting}} & NSGA-II & 
\textbf{$<$.01}/{\textbf{0.005}} & \textbf{$<$.01}/\textbf{0.005} & \textbf{$<$.01}/\textbf{0.001} & \textbf{$<$.01}/{\textbf{0.106}} \\

& \multicolumn{1}{c}{} & RS & 
\textbf{$<$.01}/{\textbf{0.082}} & \textbf{$<$.01}/\textbf{0.042} & \textbf{$<$.01}/\textbf{0.040} & \textbf{$<$.01}/{\textbf{0.179}} \\
\hline
\end{tabular}%

\begin{tablenotes}
\item[*] {*A p-value$<$0.01 indicates a significant difference between NSGA-II and RS (or \safeFuzzer); The Vargha and Delaney effect size magnitude of \Atwelve is divided into four levels, following~\cite{kitchenham2017robust}: negligible (\Atwelve $\in (0.444, 0.556)$ ), small (\Atwelve $\in (0.362, 0.444]$ ), medium (\Atwelve $\in (0.286, 0.362]$ ), large (\Atwelve $\in [0, 0.286]$ ); For consistent interpretation, \textbf{we report either \Atwelve or 1-\Atwelve, with a smaller value indicating a higher probability of NSGA-II (or RS or \safeFuzzer) better than \OriVehicleCharacteristicsSetting}; \Atwelve values at medium and large magnitudes and p-values less than 0.01 are in bold.}
\end{tablenotes}

\end{table*}

\begin{table*}[!tb]
\centering 
\caption{{Average values of the safety metrics achieved by NSGA-II/RS/\safeFuzzer/\OriVehicleCharacteristicsSetting
} -- RQ1}
\label{tab:AverageValuesOfMetric}

\renewcommand{\arraystretch}{1.5}
\begin{tabular}{cc|c|c|c}
\toprule
% \multicolumn{1}{c}{\multirow{1}[0]{*}{Case}}\\
%\cline{2-5}
Case & \safetyDegree & \tet & \tit & \aveDece \\
%& NSGA-II/RS/\OriVehicleCharacteristicsSetting & NSGA-II/RS/\OriVehicleCharacteristicsSetting & NSGA-II/RS/\OriVehicleCharacteristicsSetting & NSGA-II/RS/\OriVehicleCharacteristicsSetting \\

\hline
\precSetCarlaSun & 1.20/1.97/2.47/2.70 & 1.57/1.46/1.30/1.20 & 0.83/0.68/0.500/0.44 & 4.39/4.83/5.26/5.97 \\
\hline
\precSetCarlaRain & 0.28/1.18/1.75/2.20 & 1.72/1.67/1.52/1.40 & 1.02/0.95/0.75/0.61 & 4.31/4.67/5.29/5.42 \\
\hline
\precSetLgsvl & 7.25/8.31/-/10.1 & 3.29/2.88/-/2.20 & 2.14/1.57/-/0.72 & 1.65/1.72/-/1.87 \\
\bottomrule
\end{tabular}%

\end{table*}%
{Recall that, for each experiment, we ran \method 30 times to accommodate the randomness in searching for \VehicleCharacteristicsSettings. For \OriVehicleCharacteristicsSetting, we only run the simulation once.} Data collected from the simulation is named ``original values''. In Table~\ref{tab:OrigVSNSGAVSRSVSFUZZER}, we report the results of the statistical tests, which compare simulation results with \OriVehicleCharacteristicsSetting and with \VehicleCharacteristicsSettings generated with NSGA-II (i.e., \paretoFront){, RS and \safeFuzzer}. 
{From Table~\ref{tab:OrigVSNSGAVSRSVSFUZZER}, one can observe that, for all three settings, regarding \safetyDegree, there is a significant difference between \OriVehicleCharacteristicsSetting and NSGA-II (or RS or \safeFuzzer) produced \VehicleCharacteristicsSettings because the p-value is less than 0.01 and the \Atwelve magnitude is \emph{large} according to the definition given by Kitchenham et al.~\cite{kitchenham2017robust} (within $[0, 0.286]$), implying that \VehicleCharacteristicsSettings produced by NSGA-II, RS, and \safeFuzzer can effectively reduce the safety of the ADS vehicles compared to \OriVehicleCharacteristicsSetting. \VehicleCharacteristicsSettings generated by NSGA-II, RS, and \safeFuzzer lead to significantly larger \tet and \tit, and smaller \aveDece, with mostly large magnitudes. Overall, with the \VehicleCharacteristicsSettings generated by NSGA-II, RS, and \safeFuzzer, the vehicles are significantly more prone to unsafe situations.
% As shown in Table~\ref{tab:percChangeOrig}, \textit{TET} and \textit{TIT} values in the screened solutions are significantly larger than in the original configuration. This shows that the screened solutions reduce the safety of the vehicles with increased exposure time to unsafe situations. 
}

{Table~\ref{tab:AverageValuesOfMetric} also reports average values. For each safety metric, all algorithms generated \VehicleCharacteristicsSettings that outperformed \OriVehicleCharacteristicsSettings. For instance, NSGA-II generated \VehicleCharacteristicsSettings achieved 0.28 \safetyDegree, much less than 2.20 achieved by \OriVehicleCharacteristicsSettings. }

\begin{tcolorbox}[size=title, colframe=white, width=1\linewidth, 
breakable,
colback=gray!20]
{
{\bf Conclusion for RQ1}:
Regarding all safety metrics, in all settings, all three algorithms (NSGA-II, RS, and \safeFuzzer) can generate \VehicleCharacteristicsSettings that lead ADS vehicles into situations more unsafe than those with \OriVehicleCharacteristicsSetting.
}
\end{tcolorbox}

% Table generated by Excel2LaTeX from sheet 'metric'
% also need to update because of unscreening solution

\subsubsection{Results for RQ2}

{
As shown in Table~\ref{tab:StatisticalOthersvsNSGAII}, when comparing NSGA-II with RS, for the three settings (i.e., \precSetCarlaSun, \precSetCarlaRain and \precSetLgsvl), the 1-\Atwelve values of IGD are 1, 1 and 0.998, respectively, with the p-values less than 0.01. This clearly shows that NSGA-II significantly outperforms RS with large magnitudes in IGD. NSGA-II also significantly outperformed \safeFuzzer with large magnitudes in \precSetCarlaSun and \lgsvl, and a small magnitude in \precSetCarlaRain.}

{
Regarding \safetyDegree, \tet, \tit and \aveDece, for all three settings, NSGA-II significantly outperformed RS and \safeFuzzer with p-values less than 0.01 and \Atwelve (or 1-\Atwelve) values range from 0.579 to 0.936. In most cases, the significance is of medium or large magnitudes. These results show that it is worth using NSGA-II to solve the problem. 
}

\begin{table*}[!tb]
\centering 
\caption{Results of the Vargha and Delaney statistics and the Mann–Whitney U test — RQ2}
\label{tab:StatisticalOthersvsNSGAII}
\renewcommand{\arraystretch}{1.5}
\begin{tabular}{ccc|c|c|c|c|c}
\toprule
\multicolumn{1}{c}{\multirow{2}[0]{*}{Case}} & \multicolumn{2}{c}{\multirow{2}[0]{*}{Algorithm}} 
& IGD & \safetyDegree & \tet & \tit & \aveDece \\
\cline{4-8}
& \multicolumn{2}{c}{} & p-value/{1-\Atwelve} & p-value/{1-\Atwelve} & p-value/\Atwelve & p-value/\Atwelve & p-value/{1-\Atwelve} \\

\hline
\multicolumn{1}{c}{\multirow{2}[0]{*}{\precSetCarlaSun}} & \multicolumn{1}{c}{\multirow{2}[0]{*}{NSGA-II}} & RS & \textbf{$<$.01}/{\textbf{1.000}} & \textbf{$<$.01}/{\textbf{0.660}} & \textbf{$<$.01}/0.624 & \textbf{$<$.01}/0.619 & \textbf{$<$.01}/{\textbf{0.688}} \\

& \multicolumn{1}{c}{} & \safeFuzzer & \textbf{$<$.01}/{\textbf{0.992}} & \textbf{$<$.01}/{\textbf{0.852}} & \textbf{$<$.01}/\textbf{0.825} & \textbf{$<$.01}/\textbf{0.808} & \textbf{$<$.01}/{\textbf{0.883}}\\
\cline{1-8}
\multicolumn{1}{c}{\multirow{2}[0]{*}{\precSetCarlaRain}} & \multicolumn{1}{c}{\multirow{2}[0]{*}{NSGA-II}} & RS & \textbf{$<$.01}/{\textbf{1.000}} & \textbf{$<$.01}/{\textbf{0.653}} & \textbf{$<$.01}/0.600 & \textbf{$<$.01}/0.597 & \textbf{$<$.01}/{\textbf{0.654}} \\

& \multicolumn{1}{c}{} & \safeFuzzer & \textbf{$<$.01}/{0.603} & \textbf{$<$.01}/{\textbf{0.871}} & \textbf{$<$.01}/\textbf{0.837} & \textbf{$<$.01}/\textbf{0.820} & \textbf{$<$.01}/{\textbf{0.936}}\\

\cline{1-8}
\precSetLgsvl & NSGA-II & RS & \textbf{$<$.01}/{\textbf{0.998}} & \textbf{$<$.01}/{\textbf{0.740}} & \textbf{$<$.01}/\textbf{0.741} & \textbf{$<$.01}/\textbf{0.780} & \textbf{$<$.01}/{0.579} \\
\bottomrule
\end{tabular}%}

\begin{tablenotes}
\item[*] {*A p-value$<$0.01 indicates a significant difference between NSGA-II and RS (or \safeFuzzer); The Vargha and Delaney effect size magnitude of \Atwelve is divided into four levels, following~\cite{kitchenham2017robust}: negligible (\Atwelve $\in (0.444, 0.556)$ ), small (\Atwelve $\in [0.556, 0.638)$ ), medium (\Atwelve $\in [0.638, 0.714)$ ), large (\Atwelve $\in [0.714, 1.0]$ ); For consistent interpretation, \textbf{we report either \Atwelve or 1-\Atwelve, with a higher value indicating a higher probability of NSGA-II being better than a baseline}; \Atwelve values at medium and large levels of magnitude and p-values less than 0.01 are in bold.}
\end{tablenotes}
\end{table*}%

\begin{tcolorbox}[size=title, colframe=white, width=1\linewidth, 
breakable,
colback=gray!20]
{\bf Conclusion for RQ2:}
\method significantly outperformed RS and \safeFuzzer regarding IGD and all safety metrics; therefore, using NSGA-II is warranted.
\end{tcolorbox}

\subsubsection{Results for RQ3}\label{subsusbsec:resRQ3}

\textbf{Data preparation.} 
For each setting, we consider the reference \safetyDegree of \OriVehicleCharacteristicsSetting as the threshold \thresholdMin, based on which we select search solutions with their \safetyDegree values lower than \thresholdMin. These solutions are identified as \paretoFrontCase. In our experiment, the reference \safetyDegree values for \precSetCarlaSun, \precSetCarlaRain and \precSetLgsvl are set as 2.7, 2.2 and 10.1, respectively. These values were selected based on the simulation results with \OriVehicleCharacteristicsSettings of the virtual vehicles. Some of the \paretoFrontCase solutions led to collisions. For non-collision cases, we further employ \textit{TET} and \textit{TIT} to measure the degree of safety degradation, with a greater \tet (or \tit) value indicating a more dangerous situation. To answer all three sub-questions of RQ3, we conduct analyses based on solutions in \paretoFrontCase.

\begin{figure*}[!tb]
\centering
\begin{subfigure}{0.33\textwidth}
\centering
\resizebox{0.77\textwidth}{!}{
\begin{tikzpicture}
\begin{axis}[
label style={font=\Large},
tick label style={font=\Large},
xlabel={\# changed characteristics ($j$)},
ylabel near ticks,
ymin=-0.5,
ymax=18,
xtick=data,
xtick={1,2,3,4,5,6,7,8,9,10,11,12},
ytick={0,2,4,6,8,10,12,14,16,18},
boxplot/draw direction = y,
xtick style = {draw=none}
]
\addplot[boxplot, mark=x] table[y={n}] {experimental_data/changedParams/carla/sun/prioSun_param1.txt};
\addplot[boxplot, mark=x] table[y={n}] {experimental_data/changedParams/carla/sun/prioSun_param2.txt};
\addplot[boxplot, mark=x] table[y={n}] {experimental_data/changedParams/carla/sun/prioSun_param3.txt};
\addplot[boxplot, mark=x] table[y={n}] {experimental_data/changedParams/carla/sun/prioSun_param4.txt};
\addplot[boxplot, mark=x] table[y={n}] {experimental_data/changedParams/carla/sun/prioSun_param5.txt};
\addplot[boxplot, mark=x] table[y={n}] {experimental_data/changedParams/carla/sun/prioSun_param6.txt};
\addplot[boxplot, mark=x] table[y={n}] {experimental_data/changedParams/carla/sun/prioSun_param7.txt};
\addplot[boxplot, mark=x] table[y={n}] {experimental_data/changedParams/carla/sun/prioSun_param8.txt};
\addplot[boxplot, mark=x] table[y={n}] {experimental_data/changedParams/carla/sun/prioSun_param9.txt};
\addplot[boxplot, mark=x] table[y={n}] {experimental_data/changedParams/carla/sun/prioSun_param10.txt};
\addplot[boxplot, mark=x] table[y={n}] {experimental_data/changedParams/carla/sun/prioSun_param11.txt};
\addplot[boxplot, mark=x] table[y={n}] {experimental_data/changedParams/carla/sun/prioSun_param12.txt};
\end{axis}
\end{tikzpicture}
}
\caption{\precSetCarlaSun}
\label{fig:numberOfChangedParametersPrec1}
\end{subfigure}
\begin{subfigure}{0.33\textwidth}
\centering
\resizebox{0.77\textwidth}{!}{
\begin{tikzpicture}
\begin{axis}[
label style={font=\Large},
tick label style={font=\Large},
xlabel={\# changed characteristics ($j$)},
ylabel near ticks,
ymin=-0.5,
ymax=18,
xtick=data,
xtick={1,2,3,4,5,6,7,8,9,10,11,12},
ytick={0,2,4,6,8,10,12,14,16,18},
boxplot/draw direction = y,
xtick style = {draw=none}
]
\addplot[boxplot, mark=x] table[y={n}] {experimental_data/changedParams/carla/rain/prioRain_param1.txt};
\addplot[boxplot, mark=x] table[y={n}] {experimental_data/changedParams/carla/rain/prioRain_param2.txt};
\addplot[boxplot, mark=x] table[y={n}] {experimental_data/changedParams/carla/rain/prioRain_param3.txt};
\addplot[boxplot, mark=x] table[y={n}] {experimental_data/changedParams/carla/rain/prioRain_param4.txt};
\addplot[boxplot, mark=x] table[y={n}] {experimental_data/changedParams/carla/rain/prioRain_param5.txt};
\addplot[boxplot, mark=x] table[y={n}] {experimental_data/changedParams/carla/rain/prioRain_param6.txt};
\addplot[boxplot, mark=x] table[y={n}] {experimental_data/changedParams/carla/rain/prioRain_param7.txt};
\addplot[boxplot, mark=x] table[y={n}] {experimental_data/changedParams/carla/rain/prioRain_param8.txt};
\addplot[boxplot, mark=x] table[y={n}] {experimental_data/changedParams/carla/rain/prioRain_param9.txt};
\addplot[boxplot, mark=x] table[y={n}] {experimental_data/changedParams/carla/rain/prioRain_param10.txt};
\addplot[boxplot, mark=x] table[y={n}] {experimental_data/changedParams/carla/rain/prioRain_param11.txt};
\addplot[boxplot, mark=x] table[y={n}] {experimental_data/changedParams/carla/rain/prioRain_param12.txt};
\end{axis}
\end{tikzpicture}
}
\caption{\precSetCarlaRain}
\label{fig:numberOfChangedParametersPrec2}
\end{subfigure}
\begin{subfigure}{0.33\textwidth}
\centering
\resizebox{0.77\textwidth}{!}{
\begin{tikzpicture}
\begin{axis}[
label style={font=\Large},
tick label style={font=\Large},
xlabel={\# changed characteristics ($j$)},
ylabel near ticks,
ymin=-0.5,
ymax=18,
xtick=data,
xtick={1,2,3,4,5,6,7,8,9,10,11,12},
ytick={0,2,4,6,8,10,12,14,16,18},
boxplot/draw direction = y,
xtick style = {draw=none}
]
\addplot[boxplot, mark=x] table[y={n}] {experimental_data/changedParams/lgsvl/priolgsvl_param1.txt};
\addplot[boxplot, mark=x] table[y={n}] {experimental_data/changedParams/lgsvl/priolgsvl_param2.txt};
\addplot[boxplot, mark=x] table[y={n}] {experimental_data/changedParams/lgsvl/priolgsvl_param3.txt};
\addplot[boxplot, mark=x] table[y={n}] {experimental_data/changedParams/lgsvl/priolgsvl_param4.txt};
\addplot[boxplot, mark=x] table[y={n}] {experimental_data/changedParams/lgsvl/priolgsvl_param5.txt};
\addplot[boxplot, mark=x] table[y={n}] {experimental_data/changedParams/lgsvl/priolgsvl_param6.txt};
\addplot[boxplot, mark=x] table[y={n}] {experimental_data/changedParams/lgsvl/priolgsvl_param7.txt};
\addplot[boxplot, mark=x] table[y={n}] {experimental_data/changedParams/lgsvl/priolgsvl_param8.txt};
\addplot[boxplot, mark=x] table[y={n}] {experimental_data/changedParams/lgsvl/priolgsvl_param9.txt};
\addplot[boxplot, mark=x] table[y={n}] {experimental_data/changedParams/lgsvl/priolgsvl_param10.txt};
\addplot[boxplot, mark=x] table[y={n}] {experimental_data/changedParams/lgsvl/priolgsvl_param11.txt};
\addplot[boxplot, mark=x] table[y={n}] {experimental_data/changedParams/lgsvl/priolgsvl_param12.txt};
\end{axis}
\end{tikzpicture}
}
\caption{\precSetLgsvl}
\label{fig:numberOfChangedParametersPrec3}
\end{subfigure}
\caption{Descriptive statistics of the counts of cases in \paretoFrontCase that have a $j$ ($j=1, \ldots, 12$) number of vehicle characteristics selected over the solutions of the 30 runs (y-axis) - RQ3.1}
\label{fig:numberOfChangedParameters}
\end{figure*}
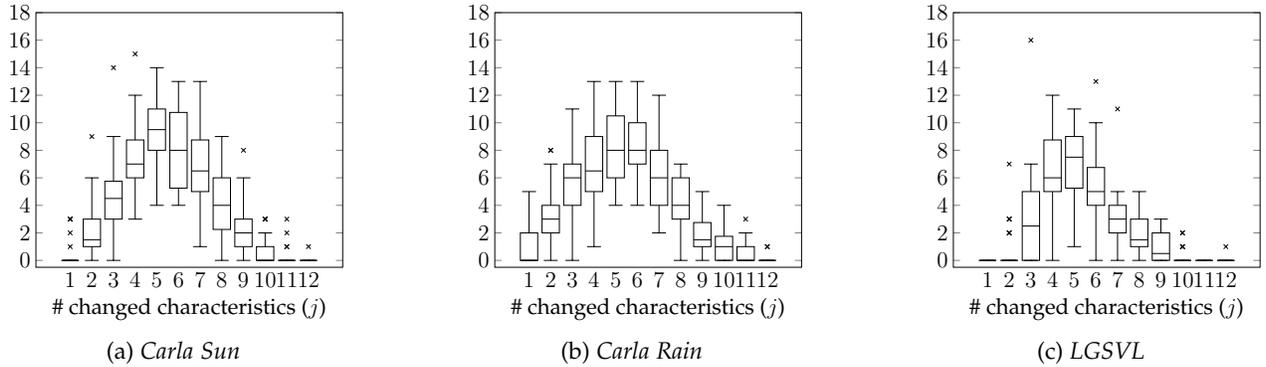

\textbf{Results for RQ3.1.}
With RQ3.1, we aim to understand to what extent the search can minimise the number of characteristics that all together negatively impact the vehicle's safety. Fig.~\ref{fig:numberOfChangedParameters} presents the descriptive statistics of the number of characteristics being changed in the search solutions: $j$ ($j = 1, \ldots, 12$) for all three settings. Note that each boxplot was plotted with \paretoFrontCase of the 30 runs of the search.

From Fig.~\ref{fig:numberOfChangedParameters}, we can observe that the number of changed characteristics for all three settings is primarily around 4, 5, and 6. For \precSetCarlaRain, the search found some cases where only one characteristic was changed. But for \precSetCarlaSun, {fewer cases changing one characteristic are found}, which are reported as outliers in Fig~\ref{fig:numberOfChangedParametersPrec1}. This finding is critical because, under poor weather conditions, changing one characteristic value might expose the vehicle to unsafe situations. {Similarly, for solutions with 2 and 3 changed characteristics, the search in \precSetCarlaRain achieved higher counts. We can also observe that the search in \precSetLgsvl finds no solutions with one characteristic change}, and solutions of the category of having two characteristics changed (denoted as \categoryOfcharacteristicsChanged$_2$ for convenience) are rare, implying that it is difficult to change just one or two characteristics to achieve a safety degradation of Apollo with the designed scenarios. In general, we observe that for all three settings, the search generated more solutions with 4-6 characteristics and fewer solutions with higher numbers of characteristics, implying that \method effectively minimises the number of changed characteristics required to harm safety.
%
%\begin{center}
%\fcolorbox{black}{gray!10}%{\parbox{.9\linewidth}{
\begin{tcolorbox}[size=title, colframe=white, width=1\linewidth, 
breakable,
colback=gray!20]
{{\bf Conclusion for RQ3.1}:}
\method {generates} \VehicleCharacteristicsSettings {with around 4–6 changed vehicle characteristics. Regarding the ability to lead the vehicle into unsafe situations, compared to} \precSetCarlaSun, in \precSetCarlaRain {it is more likely to generate} \VehicleCharacteristicsSettings {with fewer characteristics changed.} %Making a safety degradation of Apollo requires more characteristics changes.
\end{tcolorbox}
%}}
%\end{center}

\textbf{Results for RQ3.2.}
%RQ3.2 concerns which characteristics are changed more often than others. Table~\ref{tab:changeAmount} shows the results. For each characteristic $C_i$, the table reports the percentage of solutions having $j$ ($j=1, \ldots, 12$) changed characteristics in which $C_i$ is changed.
% Table generated by Excel2LaTeX from sheet 'carla_sun_out_percentage_2.7'
\begin{table}[!tb]
\caption{Percentage (\%) of occurrences of the characteristics in solutions of \categoryOfcharacteristicsChanged$_j$ ($j=1, \ldots, 12$) - RQ3.2}
\label{tab:changeAmount}
\begin{subtable}{1\columnwidth}
\centering
\caption{\precSetCarlaSun}
\setlength{\tabcolsep}{2pt}
\label{tab:changeAmountCarlaSun}
\resizebox{\textwidth}{!}{
\begin{tabular}{ccccccccccccccc}
\toprule
characteristic & \multicolumn{13}{c}{\# Changed characteristics ($j$)} & Rank\\
\cline{2-14}
& 1 & 2 & 3 & 4 & 5 & 6 & 7 & 8 & 9 & 10 & 11 & 12 & any\\
\midrule
\maxRpm & 0 & 6 & 16 & 37 & 43 & 61 & 71 & 82 & 89 & 100 & 100 & 100 & 52 & 6\\
\dampingRateFullThrottle & 0 & 0 & 2 & 1 & 5 & 9 & 20 & 38 & 56 & 77 & 86 & 100 & 14 & 10\\
\dampingRateZeroThrottleClutchEngaged & 8 & 21 & 23 & 24 & 31 & 44 & 72 & 83 & 97 & 100 & 100 & 100 & 46 & 7\\
\dampingRateZeroThrottleClutchDisengaged & 0 & 0 & 1 & 3 & 5 & 11 & 19 & 30 & 73 & 82 & 100 & 100 & 15 & 9\\
\gearSwitchTime & 8 & 15 & 24 & 32 & 68 & 86 & 93 & 96 & 92 & 95 & 100 & 100 & 66 & 4 \\
\clutchStrength & 0 & 0 & 3 & 14 & 14 & 26 & 36 & 44 & 61 & 95 & 86 & 100 & 24 & 8 \\
\mass & 17 & 30 & 64 & 78 & 88 & 91 & 96 & 98 & 100 & 100 & 100 & 100 & 84 & 3\\
\dragCoefficient & 0 & 0 & 0 & 1 & 0 & 6 & 7 & 18 & 20 & 32 & 71 & 100 & 6 & 12\\
\tireFriction & 33 & 11 & 9 & 32 & 49 & 62 & 75 & 88 & 86 & 82 & 100 & 100 & 53 & 5\\
\dampingRate & 0 & 0 & 0 & 1 & 3 & 7 & 13 & 24 & 28 & 36 & 57 & 100 & 9 & 11\\
\radius & 0 & 44 & 72 & 89 & 96 & 98 & 98 & 99 & 98 & 100 & 100 & 100 & 90 & 2\\
\maxBrakeTorque & 33 & 73 & 86 & 88 & 96 & 99 & 99 & 98 & 98 & 100 & 100 & 100 & 93 & 1 \\
\bottomrule
\end{tabular}%
}
\end{subtable}
\vspace{15pt}
\begin{subtable}{1\columnwidth}
\centering
\caption{\precSetCarlaRain}
\label{tab:changeAmountCarlaRain}
\setlength{\tabcolsep}{2pt}
\resizebox{\textwidth}{!}{
\begin{tabular}{ccccccccccccccc}
\toprule
characteristic & \multicolumn{13}{c}{\# Changed characteristics ($j$)} & Rank\\
\cline{2-14}
& 1 & 2 & 3 & 4 & 5 & 6 & 7 & 8 & 9 & 10 & 11 & 12 & any \\
\midrule
\maxRpm & 24 & 38 & 52 & 59 & 69 & 83 & 87 & 94 & 100 & 100 & 100 & 100 & 71 & 3 \\
\dampingRateFullThrottle & 0 & 0 & 0 & 1 & 4 & 10 & 27 & 39 & 54 & 70 & 93 & 100 & 14 & 9 \\
\dampingRateZeroThrottleClutchEngaged & 10 & 14 & 44 & 60 & 76 & 87 & 90 & 95 & 98 & 97 & 100 & 100 & 71 & 3 \\
\dampingRateZeroThrottleClutchDisengaged & 0 & 2 & 0 & 3 & 4 & 14 & 28 & 34 & 43 & 43 & 60 & 100 & 14 & 9 \\
\gearSwitchTime & 0 & 3 & 1 & 6 & 13 & 21 & 38 & 53 & 65 & 80 & 93 & 100 & 22 & 8 \\
\clutchStrength & 0 & 2 & 7 & 12 & 15 & 28 & 45 & 66 & 74 & 77 & 80 & 100 & 27 & 7 \\
\mass & 37 & 52 & 77 & 82 & 90 & 95 & 98 & 99 & 100 & 100 & 100 & 100 & 86 & 2 \\
\dragCoefficient & 0 & 0 & 0 & 1 & 2 & 7 & 17 & 29 & 41 & 60 & 73 & 100 & 10 & 12 \\
\tireFriction & 0 & 3 & 7 & 31 & 53 & 64 & 66 & 72 & 80 & 83 & 100 & 100 & 47 & 6 \\
\dampingRate & 0 & 1 & 1 & 3 & 4 & 8 & 16 & 25 & 54 & 93 & 100 & 100 & 12 & 11 \\
\radius & 3 & 22 & 23 & 56 & 72 & 85 & 89 & 94 & 93 & 97 & 100 & 100 & 66 & 5 \\
\maxBrakeTorque & 26 & 63 & 88 & 87 & 98 & 98 & 99 & 100 & 100 & 100 & 100 & 100 & 91 & 1 \\
\bottomrule
\end{tabular}%
}
\end{subtable}
\vspace{15pt}
\begin{subtable}{1\columnwidth}
\centering
\caption{\precSetLgsvl}
\label{tab:changeAmountLGSVL}
\setlength{\tabcolsep}{2pt}
\resizebox{\textwidth}{!}{
\begin{tabular}{ccccccccccccccc}
\toprule
characteristic & \multicolumn{13}{c}{\# Changed characteristics ($j$)} & Rank\\
\cline{2-14}
& 1 & 2 & 3 & 4 & 5 & 6 & 7 & 8 & 9 & 10 & 11 & 12 & any \\
\midrule
\maxRpm & 0 & 14 & 19 & 34 & 53 & 70 & 81 & 84 & 96 & 100 & 0 & 100 & 54 & 4 \\
\wheelMass & 0 & 0 & 0 & 14 & 41 & 59 & 67 & 73 & 74 & 86 & 0 & 100 & 39 & 6 \\
\shiftTime & 0 & 0 & 0 & 1.1 & 3 & 5 & 8 & 11 & 26 & 71 & 0 & 100 & 5 & 12 \\
\tractionControlSlipLimit & 0 & 0 & 0 & 4 & 14 & 26 & 40 & 49 & 67 & 71 & 0 & 100 & 19 & 8 \\
\minRpm & 0 & 0 & 6 & 11 & 21 & 38 & 56 & 67 & 78 & 57 & 0 & 100 & 29 & 7 \\
\maxMotorTorque & 0 & 9 & 23 & 32 & 47 & 60 & 62 & 87 & 93 & 100 & 0 & 100 & 49 & 5 \\
\maxSteeringAngle & 0 & 0 & 0 & 1.6 & 11 & 18 & 34 & 47 & 56 & 100 & 0 & 100 & 16 & 10 \\
\mass & 0 & 86 & 84 & 98 & 100 & 100 & 100 & 100 & 100 & 100 & 0 & 100 & 97 & 2 \\
\tireDragCoeff & 0 & 0 & 0 & 11 & 12 & 18 & 34 & 34 & 48 & 57 & 0 & 100 & 17 & 9 \\
\wheelDamping & 0 & 0 & 0 & 0 & 3 & 8 & 19 & 47 & 63 & 57 & 0 & 100 & 10 & 11 \\
\radius & 0 & 0 & 69 & 94 & 96 & 99 & 100 & 100 & 100 & 100 & 0 & 100 & 92 & 3 \\
\maxBrakeTorque & 0 & 91 & 100 & 99 & 100 & 99 & 99 & 100 & 100 & 100 & 0 & 100 & 99 & 1 \\
\bottomrule
\end{tabular}%
}
\end{subtable}
\end{table}%
%
%
% \begin{figure*}
% \centering
% \begin{subfigure}{1\textwidth}
% \centering
% \includegraphics[scale=0.2]{images/draw_3.2_sun.png}
% \caption{Case \precSetCarlaSun}
% \label{fig:PercentSun}
% \end{subfigure}
% \quad
% \begin{subfigure}{1\textwidth}
% \centering
% \includegraphics[scale=0.2]{images/draw_3.2_rain.png}
% \caption{Case \precSetCarlaRain}
% \label{fig:PercentRain}
% \end{subfigure}
% \caption{Percentage (\%) of occurrences of the parameters in solutions having $j$ changed parameters ($j=1, \ldots, 12$)}
% \label{fig:PrecentPic}
% \end{figure*}
RQ3.2 concerns which characteristics are changed more often than others. Table~\ref{tab:changeAmount} shows results, in which, for each characteristic $C_i$, the percentage of solutions having $j$ ($j=1, \ldots, 12$) changed characteristics are reported.
For instance, in column ``3'' (\categoryOfcharacteristicsChanged$_3$) of Table~\ref{tab:changeAmountCarlaSun} (i.e., \textit{Carla Sun}), characteristics \maxBrakeTorque, \radius, and \mass were selected and changed in, respectively, 86\%, 72\% and 64\% of the solutions of \categoryOfcharacteristicsChanged$_3$ and led to more unsafe situations than \OriVehicleCharacteristicsSetting. Moreover, we notice that they are the top three characteristics that contribute to the observed decrease in the safety degree of the vehicle. The \textit{any} columns of the three tables (Table~\ref{tab:changeAmountCarlaSun}, Table~\ref{tab:changeAmountCarlaRain}, and Table~\ref{tab:changeAmountLGSVL}) report results across all the numbers of changed characteristics, and column \textit{Rank} reports the rank of the characteristics according to the corresponding value of column ``any''. For example, for \precSetCarlaSun, the top three ranked characteristics are \maxBrakeTorque, \radius, and \mass, with 93\%, 90\%, and 84\% of the solutions in which they are changed. Likewise, for \precSetCarlaRain, the top three selected and changed characteristics are \maxBrakeTorque, \mass, and \maxRpm (tie with \dampingRateZeroThrottleClutchEngaged) with 91\%, 86\%, and 71\% of the solutions in which they are changed. There is, in addition, for \precSetLgsvl, the top three characteristics are the same as for \precSetCarlaSun (although in a different order).

%\fcolorbox{black}{gray!10}%{\parbox{.9\linewidth}{
\begin{tcolorbox}[size=title, colframe=white, width=1\linewidth, 
breakable,
colback=gray!20]
{{\bf Conclusion for RQ3.2}:
We observe that \mass and \maxBrakeTorque are the most critical characteristics, implying that they should be carefully considered when designing and testing ADS vehicles.}
\end{tcolorbox}
%}}
%\end{center}

\textbf{Results for RQ3.3.}
We analyse characteristic value changes that led to the degradation of vehicle safety {and present a portion of the results of \precSetCarlaSun in Table~\ref{tab:percChangeSun}. The complete results are provided in the online repository.} %for 
For instance, let us consider \mass in solutions of \categoryOfcharacteristicsChanged$_3$ (row C7, columns 4-5). The average percentage change of \mass is 10\% in \precSetCarlaSun. 
Besides, we identify with \valueDiff the relative difference ($v_i - v_i^{\prime\prime}$), which reports the direction of the change (i.e., positive or negative) and is measured by their respective units. For instance, a 10\% positive change of \mass from its original value is 240.8 kg (row C7, \categoryOfcharacteristicsChanged$_3$), roughly the weight of four adults, while for \radius (row C11, \categoryOfcharacteristicsChanged$_2$), a 9\% decrement to its \OriVehicleCharacteristicsSetting value (0.35 cm) means a change of 30.9 mm. 

Regarding \aveSafetyDegreeD, for instance, compared with \thresholdMin (the reference \safetyDegree in \precSetCarlaSun), the average distance reduction between the vehicle and the pedestrian, when the vehicle is stopped, is 0.32m for \categoryOfcharacteristicsChanged$_2$, which is 12\% of change to \thresholdMin.
When looking at \categoryOfcharacteristicsChanged$_2$, \categoryOfcharacteristicsChanged$_3$ and \categoryOfcharacteristicsChanged$_4$, the average absolute changes of the collision velocity (row {\aveSafetyDegreeCS}) is none (implying no collision occurred), none and -1.67m/s. This illustrates the importance of studying the influence of characteristic interactions on the safety of ADSs. Our approach produces such interactions, manifested as varying numbers of selected characteristics and their values, so that engineers can analyse or test the design of ADSs of interest based on such results.
Also, the increase of \textit{TET} and \textit{TIT} means an increase in the exposure time to dangerous situations, indicating a solution generated by the search leads to a decline in vehicle safety. 
Furthermore, regarding \aveDece, we found that the average deceleration in the solutions generated by the search is lower than that of \OriVehicleCharacteristicsSetting, which, to a certain extent, quantitatively reflects that \method generated \VehicleCharacteristicsSettings affect the control of the ADS vehicle braking system. 

% Table generated by Excel2LaTeX from sheet 
\begin{table}[!tb]
\caption{Average changes of the values of the characteristics for \categoryOfcharacteristicsChanged$_i$ {($i$ = 2, 3, \ldots, 6)} across all 30 runs w.r.t. the original value, as percentage change (\percChange = $|v_i - v_i^{\prime\prime}|/{v_i}$) and relative difference (\valueDiff = $v_i - v_i^{\prime\prime}$), in solutions of \categoryOfcharacteristicsChanged$_j$ ($j=1, \ldots, 12$) {-- \precSetCarlaSun} -- RQ3.3. }
\label{tab:percChangeSun}
% \begin{subtable}{1\textwidth}
% \centering
\resizebox{1\columnwidth}{!}{
\begin{tabular}{lllllllllll}
\toprule
Chara. (unit) & \multicolumn{10}{c}{\# Changed characteristics ($j$)}\\
\cline{2-11}
& \multicolumn{2}{c}{2} & \multicolumn{2}{c}{3} & \multicolumn{2}{c}{4} & \multicolumn{2}{c}{5} & \multicolumn{2}{c}{6} \\
& \percChange & \valueDiff & \percChange & \valueDiff & \percChange & \valueDiff & \percChange & \valueDiff & \percChange & \valueDiff \\
\midrule
\multirow{2}[0]{*}{C1 (r/min)} & 1 & 80.52 & 6 & 329.66 & 5 & 287.28 & 5 & 272.27 & 4 & 246.4\\
& 0 & 0 & 8 & -460.38 & 9 & -541.85 & 11 & -617.11 & 14 & -785.12 \\
\hline
\multirow{2}[0]{*}{C2 (kg*{m$^2$}/s)} & 0 & 0 & 7 & 0.01 & 7 & 0.01 & 13 & 0.02 & 20 & 0.03 \\
& 0 & 0 & 7 & -0.01 & 7 & -0.01 & 7 & -0.01 & 13 & -0.02 \\
\hline
\multirow{2}[0]{*}{C3 (kg*{m$^2$}/s)} & 6 & 0.11 & 6 & 0.13 & 6 & 0.12 & 8 & 0.16 & 7 & 0.14 \\
& 26 & -0.53 & 26 & -0.52 & 36 & -0.71 & 34 & -0.69 & 32 & -0.65 \\
\hline
\multirow{2}[0]{*}{C4 (kg*{m$^2$}/s) } & 0 & 0 & 11 & 0.04 & 6 & 0.02 & 9 & 0.03 & 9 & 0.03 \\
& 0 & 0 & 0 & 0 & 11 & -0.04 & 9 & -0.03 & 17 & -0.06 \\
\hline
\multirow{2}[0]{*}{C5 (s)} & 0 & 0 & 18 & 0.09 & 14 & 0.07 & 16 & 0.08 & 16 & 0.08 \\
& 16 & -0.08 & 14 & -0.07 & 16 & -0.08 & 16 & -0.08 & 20 & -0.1 \\
\hline
\multirow{2}[0]{*}{C6 (kg*{m$^2$}/s)} & 0 & 0 & 5 & 0.49 & 6 & 0.56 & 7 & 0.7 & 7 & 0.67 \\
& 0 & 0 & 2 & -0.16 & 4 & -0.39 & 5 & -0.51 & 6 & -0.57 \\
\hline
\multirow{2}[0]{*}{C7 (kg)} & 9 & 204.74 & 10 & 240.8 & 10 & 237.58 & 10 & 238.37 & 10 & 238.43 \\
& 0 & 0 & 0 & 0 & 0 & 0 & 1 & -15.53 & 1 & -27.71 \\
\hline
\multirow{2}[0]{*}{C8 (-)} & 0 & 0 & 0 & 0 & 0 & 0 & 0 & 0 & 13 & 0.04 \\
& 0 & 0 & 0 & 0 & 10 & -0.03 & 0 & 0 & 13 & -0.04 \\
\hline
\multirow{2}[0]{*}{C9 (-)} & 0 & 0 & 5 & 0.17 & 4 & 0.13 & 7 & 0.26 & 7 & 0.23 \\
& 68 & -2.39 & 59 & -2.05 & 56 & -1.97 & 58 & -2.04 & 55 & -1.92 \\
\hline
\multirow{2}[0]{*}{C10 (kg*{m$^2$}/s)} & 0 & 0 & 0 & 0 & 12 & 0.03 & 4 & 0.01 & 8 & 0.02 \\
& 0 & 0 & 0 & 0 & 4 & -0.01 & 0 & 0 & 8 & -0.02 \\
\hline
\multirow{2}[0]{*}{C11 (cm)} & 0 & 0 & 3 & 1.23 & 2 & 0.57 & 1 & 0.28 & 1 & 0.29 \\
& 9 & -3.09 & 9 & -3.28 & 9 & -3.24 & 9 & -3.2 & 9 & -3.26 \\
\hline
\multirow{2}[0]{*}{C12 (N*m)} & 2 & 31.16 & 0 & 0 & 4 & 60.47 & 4 & 65.6 & 2 & 29.94 \\
& 16 & -241.41 & 14 & -213.68 & 15 & -225.2 & 16 & -232.65 & 16 & -243.17 \\
\midrule
% \safetyDegree(-) & 12.0 & 0.32 & 24.3 & 0.66 & 37.1 & 1.00 & 55.7 & 1.50 & 76.5 & 2.06 & 91.6 & 2.47 & 87.5 & 2.36 & 78.2 & 2.11 & 90.3 & 2.44 & 84.4 & 2.28 & 64.7 & 1.75 & 60.6 & 1.64 \\
\aveSafetyDegreeD($m$) & 12.0 & 0.32 & 24.3 & 0.66 & 34.8 & 0.94 & 46.8 & 1.26 & 52.2 & 1.41 \\
\aveSafetyDegreeCS($m/s$) & - & - & - & - & - & -1.67 & - & -2.27 & - & -2.36 \\
\tet(s) & 11.9 & 0.14 & 20.5 & 0.25 & 26.8 & 0.32 & 36.8 & 0.44 & 41.1 & 0.49 \\
\tit($s^2$) & 12.3 & 0.06 & 37 & 0.17 & 65.2 & 0.29 & 105 & 0.47 & 121 & 0.54 \\
\aveDece ($m/s^2$) & 16.7 & 1.0 & 21.5 & 1.29 & 23.9 & 1.42 & 28.1 & 1.68 & 31.8 & 1.90 \\
\bottomrule
\end{tabular}%
}
\begin{flushleft} 
{{\footnotesize*C1: \maxRpm; C2: \dampingRateFullThrottle; C3: \dampingRateZeroThrottleClutchEngaged;\\ C4: \dampingRateZeroThrottleClutchDisengaged; C5: \gearSwitchTime; \\C6: \clutchStrength; C7: \mass; C8: \dragCoefficient; C9: \tireFriction; C10: \dampingRate; C11: \radius; and C12: \maxBrakeTorque.
*\aveSafetyDegreeD denotes the average reduction of distance to an obstacle: $(\sum_{i=1}^{n}{\thresholdMin - {\safetyDegree_i}})/n$ when no collision occurs; \aveSafetyDegreeCS represents the average collision velocity when a collision occurs. 
}
}
\end{flushleft}

\end{table}

%\begin{center}
%\fcolorbox{black}{gray!10}%{\parbox{.9\linewidth}{
\begin{tcolorbox}[size=title, colframe=white, width=1\linewidth, 
breakable,
colback=gray!20]
{\textbf{Conclusion for RQ3.3:} Detailed results on value changes to each characteristic are valuable in conducting guided analyses and testing.}
\end{tcolorbox}
%}}
%\end{center}

\section{Discussion}\label{sec:discussion}
% \IEEEraisesectionheading{\section{DISCUSSION}\label{sec:DISCUSSION}}

\subsection{Correlation between \VehicleCharacteristicsSettings and weather conditions} 
% The results for answering RQ3 are collected from Tables~\ref{tab:changeAmountCarlaSun}, \ref{tab:percChangeOrigPrec1}, \ref{tab:changeAmountCarlaRain} , \ref{tab:percChangeOrigPrec2} and Figure~\ref{fig:CompAveChange}. 
Among the most frequently changed characteristics, \mass and \maxBrakeTorque were among the top three for all the experiments with \carla (observed in RQ3.2). We might consider that studying these two characteristics' impact on autonomous vehicles' safety is essential. On the other hand, \maxRpm and \dampingRateZeroThrottleClutchEngaged were ranked in the 3rd position in \precSetCarlaRain, but at the 6th and 7th positions for \precSetCarlaSun. 
When looking at the results from RQ3.2 (Table~\ref{tab:changeAmountCarlaSun} and \ref{tab:changeAmountCarlaRain}), we further observed that the percentages of the above two characteristics appearing in all solutions (considering all the numbers of changed characteristics) range from 52\% and 46\% (for \precSetCarlaSun) to 71\% and 71\% (for \precSetCarlaRain). Furthermore, when looking at \gearSwitchTime, the percentage of it being changed was reduced by 44\% (i.e., 66\% with \precSetCarlaSun and 22\% with \precSetCarlaRain in Table~\ref{tab:changeAmountCarlaSun} and \ref{tab:changeAmountCarlaRain}). These results show that there seem to be correlations between which vehicle characteristics are being selected by the search and different weather conditions.

%When looking at the 6th and 7th columns of both Table~\ref{tab:percChangeOrigPrec1} and \ref{tab:percChangeOrigPrec2}, 
From the results, we observed that for \VehicleCharacteristicsSettings with three characteristics changed, no collision was observed when driving on a sunny day, and the distance (\aveSafetyDegreeD ($m$)) between \car and the pedestrian decreased 24.3\% from its original \VehicleCharacteristicsSetting (\thresholdMin in \precSetCarlaSun), meaning a decrease of the distance (\aveSafetyDegreeD ($m$)) by 0.66m. In \precSetCarlaRain, for \aveSafetyDegreeD ($m$), a 30.9\% change, with regard to its original \VehicleCharacteristicsSetting, means that the distance was shortened by 0.68m on average. Furthermore, %when looking at the last row of the 7th column in Table~\ref{tab:percChangeOrigPrec2}, collisions occurred, and 
the average collision speed was 1.71m/s, which is, however, not the case for \precSetCarlaSun. Moreover, in all solutions \paretoFrontCase, %as shown in Table~\ref{tab:collisionCase}, 
we observe that more collisions occurred in setting \precSetCarlaRain; in \precSetCarlaSun, there were 157 (out of the 1486 total solutions) collisions, while 360 collisions (out of 1431 solutions) in \precSetCarlaRain. 
These observations might hint that under poorer weather conditions, changes to specific vehicle characteristics might cause a more severe safety impact on ADSs.

% Table generated by Excel2LaTeX from sheet 'carla_sun_col'

\subsection{Interaction effects of configurable characteristics}
To study the interaction effects of the characteristics, we summarise the characteristic selections of the solutions of \categoryOfcharacteristicsChanged$_4$ in Table~\ref{tab:differentCombination}, as an example. For each characteristic $C_i$, the top three combinations with the most occurrences in solutions of \categoryOfcharacteristicsChanged$_4$ are presented, and we also reported the percentage of occurrences for each combination in the last row.
\begin{table}[!tb]
\caption{The top three characteristic combinations in solutions with four changed characteristics (\categoryOfcharacteristicsChanged$_4$) (summarised in the three columns of each setting). {Note that \textit{Mass} and \textit{maxBrakeTorque} are the two characteristics selected by all the top three combinations and hence highlighted in bold.}}
\label{tab:differentCombination}
\begin{subtable}{1\columnwidth}
\centering
\caption{\precSetCarlaSun and \precSetCarlaRain}
\setlength{\tabcolsep}{2pt}
\label{tab:differentCombinationCarlaSun}

\begin{tabular}{c|ccc|ccc}
\hline
characteristic & \multicolumn{3}{c|}{\precSetCarlaSun} & \multicolumn{3}{c}{\precSetCarlaRain} \\
\hline
\maxRpm & \checkmark & $\times$ & $\times$ & \checkmark & \checkmark & $\times$ \\
\dampingRateFullThrottle & $\times$ & $\times$ & $\times$ & $\times$ & $\times$ & $\times$ \\
\dampingRateZeroThrottleClutchEngaged & $\times$ & $\times$ & $\times$ & $\times$ & \checkmark & \checkmark \\
\dampingRateZeroThrottleClutchDisengaged & $\times$ & $\times$ & $\times$ & $\times$ & $\times$ & $\times$ \\
\gearSwitchTime & $\times$ & $\times$ & $\times$ & $\times$ & $\times$ & $\times$ \\
\clutchStrength & $\times$ & $\times$ & \checkmark & $\times$ & $\times$ & $\times$ \\
\textbf{\mass} & \checkmark & \checkmark & \checkmark & \checkmark & \checkmark & \checkmark \\
\dragCoefficient & $\times$ & $\times$ & $\times$ & $\times$ & $\times$ & $\times$ \\
\tireFriction & $\times$ & \checkmark & $\times$ & $\times$ & $\times$ & $\times$ \\
\dampingRate & $\times$ & $\times$ & $\times$ & $\times$ & $\times$ & $\times$ \\
\radius & \checkmark & \checkmark & \checkmark & \checkmark & $\times$ & \checkmark \\
\textbf{\maxBrakeTorque} & \checkmark & \checkmark & \checkmark & \checkmark & \checkmark & \checkmark \\
\hline
selected\_pc & 24.4 & 16.7 & 8.1 & 22.3 & 11.7 & 9.2 \\
\hline
\end{tabular}%
\end{subtable}
\vspace{15pt}
\begin{subtable}{1\columnwidth}
\centering
\caption{\precSetLgsvl}
\label{tab:differentCombinationLGSVL}
\setlength{\tabcolsep}{2pt}

\begin{tabular}{c|ccc}
\hline
characteristic & \multicolumn{3}{c}{\precSetLgsvl} \\
\hline
\textbf{\mass} & \checkmark & \checkmark &\checkmark \\
\wheelMass & $\times$ & $\times$ & \checkmark \\
\radius & \checkmark &\checkmark &\checkmark \\
\maxRpm & $\times$ & \checkmark & $\times$ \\
\minRpm & $\times$ & $\times$ & $\times$ \\
\textbf{\maxBrakeTorque} & \checkmark & \checkmark & \checkmark \\
\maxMotorTorque & \checkmark & $\times$ & $\times$ \\
\maxSteeringAngle & $\times$ & $\times$ & $\times$ \\
\tireDragCoeff & $\times$ & $\times$ & $\times$ \\
\wheelDamping & $\times$ & $\times$ & $\times$ \\
\shiftTime & $\times$ & $\times$ & $\times$ \\
\tractionControlSlipLimit & $\times$ & $\times$ & $\times$ \\
\hline
selected\_pc & 29.9 & 27.2 & 14.1 \\
\hline
\end{tabular}%

\end{subtable}
\end{table}%
The two tables show that combinations of \mass, \radius, \maxRpm, and \maxBrakeTorque appear in all three settings, and the percentages of occurrences are all over 20\%. 
%From this observation, once again, we can conclude that when the weather conditions of the driving scenario change, the combinations of configurations of selected parameters found by the search are similar. 
This implies that, for both simulators, the combination of critical characteristics is similar for the driving scenarios designed to require \car to perform emergency braking when facing obstacles, although with different vehicle dynamics models. 
Based on this observation, our approach seems to provide useful information such that engineers can focus on combinations of characteristics frequently appearing in \VehicleCharacteristicsSettings solutions returned by the search.

%\subsection{Practical Relevance}
\subsection{{Cost-Effectiveness of \method}}
{In the industrial production process, various parts of vehicles are very complicated. For instance, among 12 parameters, up to 792 combinations can be obtained by selecting 5 out of 12 parameters. In reality, we cannot conduct production testing with many combinations; therefore, it is difficult to produce vehicles of different configuration combinations for testing within limited resources. Therefore, our proposed testing method can help automotive engineers narrow down the search scope and hence reduce the cost.}

{When evaluating the computational costs of applying \method, the most significant contribution comes from the simulations run in the simulator. In contrast, the time consumed by the NSGA-II optimization process is negligible.}

\subsection{{Reality Gap and Implications}} 
{Most ADS testing methods are simulation-based, but no simulator can fully reproduce real-world driving scenarios~\cite{khatiri2023simulation}. Whether test results can be generalized to the real world is still an open issue~\cite{AfzalICST2021}. \method also faces such problems as differences between vehicle dynamics models and real vehicle dynamics  exist. Stocco et al.~\cite{stocco2022mind} also found through an empirical study that driving in physical cars is affected by inevitable real-time randomness, which is insignificant in the virtual world. For example, surface friction and battery voltage may adversely affect the throttle, while steering angle prediction may be affected by sudden spikes in brightness, delays between prediction and activation, and other factors that may only occur in the physical environment. Theoretically, these factors could all be regarded as part of \VehicleCharacteristicsSettings and studied; however, they are very hard to simulate in today's simulators. In any case, \method is general in that, with the advance of simulators, \VehicleCharacteristicsSettings can be expanded to include more vehicle characteristics that can be configured, and their effects can be simulated.}

\section{Threats To Validity}\label{sec:threats}
% \IEEEraisesectionheading{\section{THREATS TO VALIDITY}\label{sec:THREATS TO VALIDITY}}

{\it Internal Validity.}
In the search-based context, an inherent problem is randomness in the search process, e.g., due to various aspects such as genetic operators and their chosen parameters. To lower the randomness effect on the obtained results from our experiments, we repeated our experiments 30 times, followed by collecting and analysing experimental data using appropriate statistical tests. In particular, we applied the Mann-Whitney U Test and the \Atwelve statistic on the experimental data obtained from thirty independent runs based on a well-established guide~\cite{Arcuri2011}. Moreover, we set the same termination criterion for NSGA-II, RS and \safeFuzzer, i.e., the number of fitness evaluations. Regarding the parameter settings for NSGA-II, we employ the default settings provided by the JMetalPy framework. Such default settings have shown good performance in many SBSE problems~\cite{Arcuri2013ParameterTO,lu2021search}. Regarding the population size of the NSGA-II algorithm in our approach, we set different values for it in three settings (i.e., 30 for \precSetLgsvl, 50 for both \precSetCarlaSun and \precSetCarlaRain). The simulation in our experiments is time-consuming, as one run takes roughly 14 hours (17s of one simulation*100 generations*30 individuals) for \precSetLgsvl. Then, we conducted a pilot study to analyse the convergence trend to determine the appropriate population sizes. To compare NSGA-II, RS and \safeFuzzer, we select the appropriate quality indicator (i.e., IGD) following the guidelines in~\cite{ALIQI}.

{Choosing threshold \thresholdI value will impact the search. %A sub-optimal selection could potentially affect search performance. 
However, systematically studying optimal \thresholdI for each characteristic within its domain $D_i$ on the search performance requires a detailed experiment, the results of which will guide the selection of \thresholdI. Conducting such experiments and developing a guide are our future work.}

In our experiments, we selected \carla and \lgsvl as the simulators for all three cases to provide driving scenarios. The two simulators provide different driving scenarios, but the identified critical scenarios are the same type, requiring the vehicle under test to operate (i.e., emergency braking) to avoid getting into a dangerous situation (e.g., collision). We used the metric \safetyDegree, which is shown to be suitable for measuring the performance of the operation of emergency braking~\cite{yin}. Considering the vehicles under test, 12 configurable characteristics provided by the two simulators were used in the experiments. Studying additional characteristics is one of our future works.

%The impact of extreme weather on driving safety cannot be ignored in practice. However, existing simulators have limitations on simulating the impact of extreme weather on vehicles under test (e.g., physical characteristics). We acknowledge that simulators rendering more realistic physical properties would enhance the reliability of conclusions that are carried out with them. However, there are no publicly available simulators that can provide a large set of realistic weather conditions. However, we found they impact the quality of data collected by the sensor (e.g., images blurred by raindrops) and then affect the control of the ADS algorithm (e.g., vision-based ADS) on the vehicle. Based on this impact, it is still meaningful for us to study the degree of impact of ADSs on vehicle control by changing \VehicleCharacteristicsSettings in extreme weather. Besides, to minimize the threat of impacts on the generalizability of the approach, we deployed the proposed approach in two open-source simulators and tested two ADSs. Both the two simulators are widely used in the training and testing of ADSs.
{\it External Validity.}
The impact of extreme weather on driving safety cannot be ignored in practice, as our results indicated. We could experiment with limited weather conditions in our experiments, and conducting a comprehensive study on the impact of a wide range of realistic weather conditions is needed. However, existing simulators have limitations on simulating the impact of extreme weather on vehicles under test (e.g., physical characteristics). We acknowledge that simulators rendering more realistic physical properties would enhance the reliability of conclusions. However, no publicly available simulators can provide a large set of realistic weather conditions. Moreover, to minimise the threat related to the generalisability of the approach to different simulators, we deployed our approach in two open-source simulators and tested two ADSs. Both simulators are widely used in the training and testing of ADSs. Nonetheless, experimentation with additional simulators and ADSs is warranted in the future. {Finally, we experimented with only one scenario instead of all scenarios provided in CARLA (i.e., 21) due to the high cost associated with simulation time and the required number of repetitions (i.e., 30) of experiments to deal with the inherent randomness of search algorithms. In the future, we intend to include more scenarios to determine whether our results are generalizable to other scenarios. }

\section{Conclusion And Future Work}\label{sec:conlusion}
% \IEEEraisesectionheading{\section{CONCLUSION AND FUTURE WORK}\label{sec:CONCLUSION AND FUTURE WORK}}
We studied the impact of vehicle characteristics (e.g., mass) variations on Autonomous Driving Systems (ADSs) safety. We formalized the problem of searching for variations in characteristics that negatively affect ADS safety as a multi-objective search problem. We adopted NSGA-II -- a commonly used algorithm in search-based software engineering, to solve the problem. We conducted experiments using two ADSs executed in two simulators. Through a comprehensive analysis of the experimental results, we report the identified critical characteristics that reduce the safety of ADSs. Furthermore, the combination of these critical characteristics, the range of their values, and the difference in vehicle characteristics variations in weather conditions are also reported, leading to the conclusion that weather conditions should be studied together with various ADSs' characteristics variations.

Our future plans are as follows. First, we will adopt more realistic vehicle dynamics models by using other simulators such as CarSim~\cite{2003CarSim}. Second, we will experiment with additional algorithms (e.g., SPEA2~\cite{2001SPEA2} and MoCell~\cite{2009MOCell}) to see if another algorithm can outperform NSGA-II. Third, in addition to studying safety, we want to study other performance aspects, such as the comfort of passengers. {Fourth, we will investigate other vehicle characteristics when different simulators are employed and identify constraints among them such that more realistic configurations can be generated. 
%Fifth, we would also like to integrate our approach with solutions for predicting test outcomes for driving scenarios before being executed such as those proposed by Huai et al.~\cite{huai2023doppelganger}, Birchler et al.~\cite{birchler2023machine}, and assessed by Nejati et al.~\cite{nejati2023reflections}. 
Fifth, since running simulations is expensive, we would also like to adopt approaches that avoid executing the scenarios, i.e., identifying tests that are unlikely to detect faults as proposed by Birchler et al.~\cite{birchler2023machine}, or using surrogate models in the fitness evaluation (see the survey by Nejati et al.~\cite{nejati2023reflections}). 
Sixth, we also plan to extend our work for fault injection to study the effect of malfunctioning mechanical components of vehicles on ADS safety, etc. Seventh, we intend to compare the two simulators with a carefully planned experiment with common characteristics to see the impact of parameter variations on safety across the two simulators.} Last, with detailed data analysis, we want to provide a tool with guidelines for engineers to study ADS robustness.

% or
%\appendix % for no appendix heading


%\appendices

%\section{}

\ifCLASSOPTIONcompsoc
\section*{Acknowledgments}
\else
\section*{Acknowledgment}
\fi

This work is supported by the National Natural Science Foundation of China under Grant No. 61872182. P. Arcaini is supported by Engineerable AI Techniques for Practical Applications of High-Quality Machine Learning-based Systems Project (Grant Number JPMJMI20B8), JST-Mirai; and also by ERATO HASUO Metamathematics for Systems Design Project (No. JPMJER1603), JST; Funding Reference number: 10.13039/501100009024 ERATO. This work is also supported by the Co-evolver project (No. 286898/F20), funded by the Research Council of Norway.

\ifCLASSOPTIONcaptionsoff
\newpage
\fi



\bibliographystyle{IEEEtranN}
\bibliography{biblio}


% 照片
% \end{IEEEbiography}


% % insert where needed to balance the two columns on the last page with
% %\newpage



%\vfill


\end{document}